\documentclass[manuscript]{aastex} 
\slugcomment{accepted by ApJ} \shorttitle{LSQ RR Lyrae Star Survey:
  Region I} \shortauthors{Zinn et al. }
\begin{document}
\title{La Silla QUEST RR Lyrae Star Survey:  Region I}
\author{R. Zinn\altaffilmark{1}, B. Horowitz\altaffilmark{2}, A. K. Vivas\altaffilmark{3}, C. Baltay\altaffilmark{2}, N. Ellman\altaffilmark{2}, E. Hadjiyska\altaffilmark{2}, D. Rabinowitz\altaffilmark{2}, and L. Miller\altaffilmark{1}}
\altaffiltext{1}{Department of Astronomy, Yale University, P.O. Box 208101, New Haven, CT 06520-8101}
\altaffiltext{2}{Department of Physics, Yale University, P.O. Box 208120, New Haven, CT 06511-8499}
\altaffiltext{3}{Centro de Investigaciones de Astronom\'ia, Apartado Postal 264, M\'erida 5101-A, Venezuela}
\begin{abstract}
A search for RR Lyrae stars (RRLS) in $\sim 840$ $ \textrm{deg}^2$ of
the sky in right ascension $150\degr - 210\degr$ and declination
$-10\degr - +10\degr $ yielded 1013 type ab and 359 type c RRLS.  This
sample is used to study the density profile of the Galactic halo, halo
substructures, and the Oosterhoff type of the halo over distances
($d_{\sun}$) from $\sim 5$ to $\sim 80$ kpc.  The halo is flattened
towards the Galactic plane, and its density profile steepens in slope
at galactocentric distances greater than $\sim 25 \textrm{kpc}$.  The
RRLS in the stellar stream from the Sagittarius dwarf spheroidal
(dSph) galaxy match well the model of Law \& Majewski for the stars
that were stripped 1.3 to 3.2 Gyr ago, but not for the ones stripped
3.2 to 5.0 Gyr ago.  Over densities are found at the locations of the
Virgo Overdensity and the Virgo Stellar Stream.  Within $1\degr$ of
1220-1, which Jerjen et al. identify as a halo substructure at
$d_{\sun} \sim 24$ kpc, there are 4 RRLS that are possibly members.
Away from substructures, the RRLS are a mixture of Oosterhoff types I
and II, but mostly OoI ($\sim 73\% $).  The accretion of galaxies
resembling in RRLS content the most massive Milky Way satellites (LMC,
SMC, For, Sgr) may explain this preponderance of OoI.  Six new RRLS
and 3 new anomalous Cepheids were found in the Sextans dSph galaxy.
\end{abstract}
\section{Introduction}
The properties of the Galactic halo and how they may be explained in
the broader context of galaxy evolution have been a major area of
research for more than 50 years.  Early investigations of the halo, which
by today's standards relied on tiny samples of halo stars and globular
clusters, concluded that the halo is very old and uniform in age and
has smooth contours of stellar density.  The evolutionary picture that
emerged from these studies, painted in the seminal paper by
\citet{eggen62} (ELS), was that the halo formed rapidly during an
early stage of the collapse of a single large gas cloud that later
became the galactic disk.  In today's terms, the ELS picture is an
example of \emph{in situ} formation, because it envisioned that the
formation of stars and globular clusters occurred within the confines
of the Milky Way (MW).

Later it was realized that in situ formation was probably not the
origin of the whole halo, in particular its outer parts, because the
ELS picture failed to explain the distribution and gradients in
metallicity and in horizontal-branch (HB) morphology of the halo
globular clusters (GCs) \citep{sz78} and the correlations between
metallicity and kinematics of local samples of metal-poor stars
\citep{norris85, carney90}.  This motivated an accretion picture in
which some halo stars and GCs formed in separate, small galaxies that
subsequently disintegrated as they merged with the MW, depositing
their already formed stars and star clusters into the halo.  This
picture has received very strong observational support in recent years
by the discovery of the Sagittarius dwarf spheroidal galaxy (dSph),
which is currently undergoing tidal disruption, the discoveries of
several stellar streams and overdensities in the halo that are
probably the remains of accreted dwarf galaxies or GCs
\citep[e.g.,][]{helmi99,ibata01a,vivas01, newberg02, grillmair06a,
  duffau06, clewley06, belokurov07b, belokurov07c, juric08,
  schlaufman09, morrison09,bonaca12b, sesar12,drake13b}, and the
discoveries of many ``ultra-faint'' dwarf galaxies (UFD) in the halo
\citep[e.g.,][]{willman05,zucker06,belokurov06b}, which along with the
previously known dSph galaxies are probably the survivors of a larger
population of similar objects that have already merged with the MW.
The existence of large stellar streams in the halos of other galaxies
\citep[e.g.,][]{shang98, ibata01b, zucker04, martinez09} most notably
in M31 where some appear to contain globular clusters as well as stars
\citep{mackey10}, provide conclusive evidence that accretion is a
widespread phenomenon.

While over the past 30 years many investigators have speculated that
the halo may be the result of both in-situ formation and
accretion, several recent studies have greatly strengthend the
evidence.  For example, from a study of the kinematics and
metallicities of thousands of stars near the main-sequence turnoff,
\citet{carollo07,carollo10} concluded that the
halo consists to two over-lapping components.  The inner halo,
galactocentric distances, $R_{gc}$, less than 10-15 kpc, is dominated
by a relatively metal rich component ($\langle\textrm{[Fe/H]}\rangle = -1.6$)
  that is flattened towards the galactic plane and has zero or
  slightly prograde rotation.  In contrast, the outer halo, which
  dominates at $R_{gc} \gtrsim 15 $ kpc, is nearly round, in retrograde
  rotation, and significantly more metal-poor ($\langle[Fe/H]\rangle =
  -2.2$).  This picture, which has spurred controversy
  \citep[see][]{schonrich11,beers12}, has been supported by recent
  investigations of blue horizontal branch (BHB) and RR Lyrae stars
  (RRLS)\citep[e.g.,][]{kinman12,hattori13}.  This dual halo may be
  explained by in-situ formation and accretion dominating,
  respectively, the inner and outer halos \citep{carollo07, carollo10}.

Important age information on the two halos is provided by the GC
systems.  When the halo GCs are divided into young halo (YH) and old
halo (OH) groups on the basis of the variation of HB morphology at
constant [Fe/H] (i.e., the $2^{nd}$ parameter effect), which is a
rough proxy for cluster age \citep[e.g.,][]{dotter11}, they are also
divided by kinematics and by spatial distribution
\citep{zinn93,dacosta95,mackey04,lee07}.  The very old ages, mean
prograde rotation and concentration at small $R_{gc}$ of the OH group
suggest that the majority of them formed in situ, probably during an
early stage of the dissipational collapse that led to the formation of
the bulge and disk and the concomitant bulge/disk population of GCs.
In contrast, the much more widely distributed and kinematically hotter
YH clusters, and in addition a small fraction of the OH clusters that
resemble YH clusters in these properties, are probably debris from
accretion events. This is certainly the case for the clusters that can
be unambiguously identified with the stellar streams from the Sgr dSph
galaxy \citep{carraro07,law10b}. The recent discovery that the YH
clusters have a planar spatial distribution that is very similar to
the one of the dSph and UDF galaxies indicates a common origin and is
additional evidence that these clusters originated in now destroyed
satellite galaxies \citep{keller12,pawlowski12}.  The wide range in
age of the YH clusters \citep[$\sim 5$ Gyrs][]{dotter11,carraro07}
suggests that a fraction of the accreted halo originated in several
slowly evolving dwarf galaxies that formed globular clusters over
extended periods of time, not unlike the Sgr and Fornax dSph galaxies
\citep{zinn93,dacosta95,mackey04,forbes10}.  However, the low mean
      [Fe/H] of the outer halo suggests that much of it originated in
      low-mass systems that never formed clusters and experienced less
      chemical evolution \citep{carollo07,carollo10}.

The first investigations of the relative abundances of the elements in
the dSph galaxies indicated that they do not resemble the metal-poor
stars near the Sun in the ratio of the $\alpha $ elements to Fe
\citep[e.g.,][]{shetrone01}, which cast doubt on the hypothesis that
satellite accretion played a role in halo formation.  As larger
samples stars in the galaxies were measured, it became clear that the
very metal-poor stars in the galaxies and in the halo had similar
abundance ratios, but not the more metal-rich ones.  Since many of the
surviving dSph galaxies have experienced long histories of star
formation and chemical enrichment, it was argued that the the
abundances of their metal-rich stars may not be indicative of the
stars in dwarf galaxies that were accreted in the early evolution of
the MW \citep[see reviews by][]{geisler07,tolstoy09}.
Most of the halo stars that were used in these comparisons with the
dSph galaxies are members of the inner halo.  Some recent
investigations of the chemical abundances of halo stars have separated
them into inner and outer halo samples on the basis of kinematics,
which has led to the discovery of important differences.
\citet{roederer09} found that the inner halo is chemically more
homogeneous than the outer halo, which can be explained if the
accretion of dwarf galaxies with different chemical enrichment
histories played a larger role in the outer than in the inner halo.  A
series of recent papers \citep{nissen10,nissen11,schuster12} has shown
that there are correlations between $\alpha$/Fe and other abundance
ratios and the kinematics and the ages of halo stars.  The stars in
the outer halo are 2-3 Gyrs younger and have lower $\alpha$/Fe than
the ones in the inner halo, again consistent with more accretion in
the outer halo.  There is evidence that satellite accretion also
contributed to the inner halo, for \citet{sheffield12} have shown that
some relatively nearby halo M giants have abundance ratios that are
unlike the majority of halo stars, but similar to the stars in
dwarf galaxies.  In a reversal of what was once the case, the
comparisons of the chemical compositions of the satellite dwarf
galaxies and halo stars provide strong evidence that the accretion of
similar galaxies contributed to the MW halo \citep[see
  also][]{frebel13}.

Clearly, great progress has been made in understanding the formation
of the halo, but there is a need for more observations.  The number
and the characteristics of the accretion events and of the surviving
satellite galaxies provide tests of the relative importance of the
accretion and in situ scenarios.  While there is universal agreement
that halo substructures exist, estimates of the fraction of the halo
contained in substructures vary substantially
\citep[e.g.,][]{bell07,xue11,deason11}.  The transition zone from
$R_{gc} \sim 10 \, \textrm{ to} \sim 30$ kpc is of particular interest,
because it is not clear if the properties of the halo vary smoothly or if
there is a fairly abrupt transition.  Because the MW is the best
observed galaxy, the observations of its halo provide important
constraints on simulations of galaxy formation in a cosmological
context \citep[e.g.,][]{bullock05,zolotov09,font11,helmi11}.  With the
goals providing additional data on halo structure and the presence and
absence of stellar streams over the range of roughly 10 to 80 kpc, we
have embarked on this series of papers.

This paper reports the first results of the La Silla -- QUEST (LSQ)
survey for RR Lyrae stars (RRLS) which has employed the 1-m Schmidt
telescope at La Silla, in Chile and the QUEST camera.  RRLS are, of
course, pulsating horizontal branch stars with very characteristic
periods ($0.2$ to $0.9$\ day) and lightcurves.  Because their
amplitudes vary between about 0.2 and 1.3 magnitudes in V, they are
easily detected in multi-epoch photometric surveys.  The most
important property of RRLS as probes of the halo is that they are
excellent standard candles and, as previous surveys have shown
\citep[e.g.,][]{ivezic00,vivas01,keller08,watkins09,sesar10,drake13b},
can be used to detect halo substructure.

In this paper, we describe our methods for selecting candidate RRLS
and measuring the periods and lightcurves for the RRLS that we have
identified in contiguous region of $\sim 840$ $\textrm{deg}^2$, spanning
the constellations Sextans to Virgo.  We use these stars to
investigate the density profile the Galactic Halo and find a
significant break at $R_{gc}\sim25$ kpc.  We also investigate the
previously identified overdensities in the region, the leading stream
from the Sgr dSph galaxy and the Virgo Stellar Stream, and examine the
data for signatures of other streams.  Finally, we investigate the
pulsational properties of the RRLS (i.e., the Oosterhoff Effect) and
compare the halo RRLS to those found in the four most massive
satellite galaxies of the MW.  We also see if there are variations in
RRLS properties across the break in the density profile of the halo.

\section{Observations}

\subsection{Instrumentation}

The RRLS survey reported here uses the data from the La Silla-QUEST
Southern Hemisphere Variability Survey \citep{hadjiyska12}, which is
primarily a survey for supernovae \citep{baltay13} and trans-Neptunian
objects.  The Survey employs the 10 square degree QUEST Camera (Baltay
et al 2007) at the prime focus of the 1 meter ESO Schmidt Telescope at
the La Silla Observatory in Chile. The survey commenced observations
in September of 2009 and uses 90\% of the telescope time.  For both
the La Silla-QUEST (LSQ) RRLS and supernovae surveys, the telescope
was used in the point-and-stare mode with 60s exposures, rather than
the drift-scan mode, which was used in the Venezuelan QUEST RRLS
survey \citep{vivas04} (hereafter, simply the QUEST survey).
Typically, each night of observation consisted of two visits to the
same position, separated by $\sim 2$ hours, which allowed the
elimination of moving objects from the data.  Within one or two days,
the observations were repeated, and the time spans over which observations at
same position were obtained ranged from $\sim25$ to $\sim90$ days.

\begin{figure}
\epsscale{0.9}
\plotone{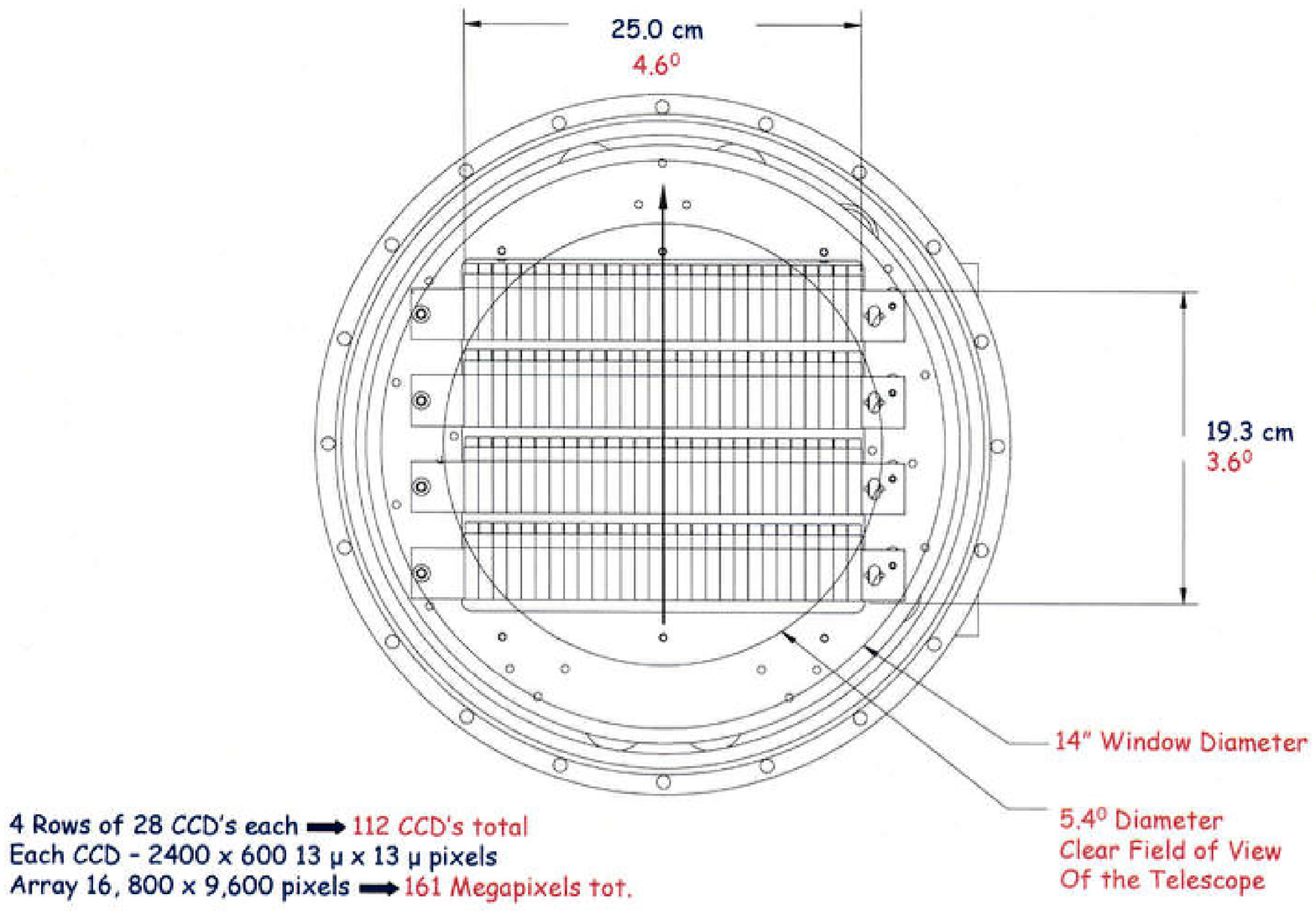}
\caption{The layout of the La Silla QUEST CCD camera.  Each of the 4
  horizontal rows contains 28 CCD's, which are depicted as narrow
  rectangles.}
\label{fig-camera}
\end{figure}

The camera consists of 112 CCD detectors arranged in a 4 by 28 array
(4 rows, labeled A, B, C, and D, with 28 CCD's each) as shown in
Figure~\ref{fig-camera}. The progression of the 28 CCDs on a row are
in the declination direction, with CCD 1 on the north side and CCD 28
on the south side of the field of view. Finger A is on the east side
and finger D is on the west side of the field.

The CCD's are 600 x 2400 pixel thinned, back illuminated SARNOFF
devices with $13\mu$m x $13 \mu$m pixels and a plate scale of 0.86
arcsec/pixel. The camera covers $3.6\degr$ x $4.6\degr$ on the
sky with an active area of 9.6 $\textrm{deg}^2$, when all 112 CCDs are
operating.  During the RRLS survey, $\sim 100$ CCDs, 89\% of the
active area was utilized.
                                                                                
\begin{figure}
\epsscale{1.0}
\plotone{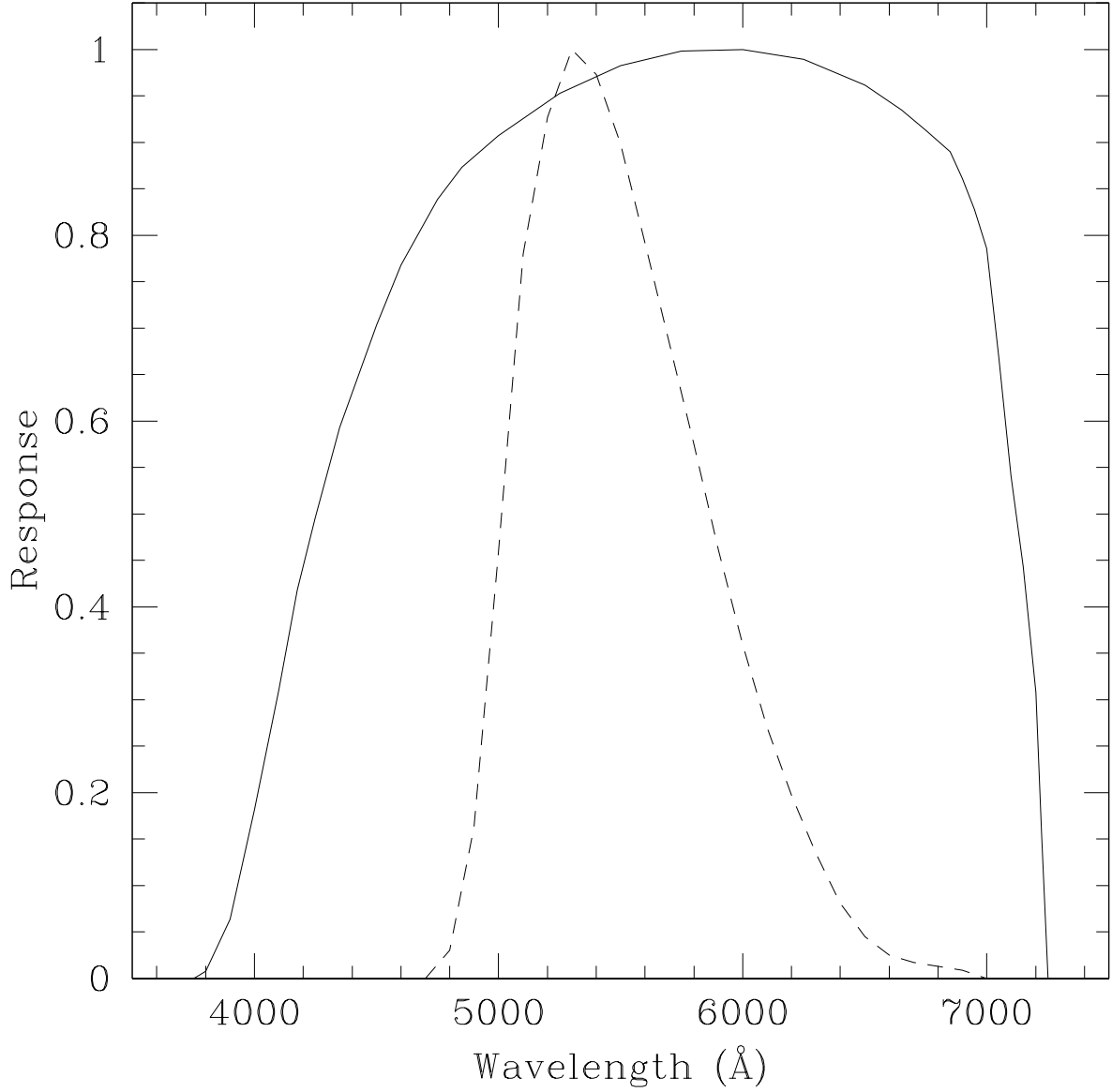}
\caption{The solid curve is the response function of the filter plus
  CCD used in the LSQ RRLS survey.  The dashed curve is the response
  function of the V passband as determined by \citet{bessell90}.}
\label{fig-filter}
\end{figure}

The survey uses a very broad-band filter to enable the detection of
faint sources in short exposures.  Plotted in Figure~\ref{fig-filter}
is the normalized response function ($S(\lambda)$) of this filter and
the camera CCD's, which was determined from the transmission curve of
the filter and the spectral sensitivity of a typical CCD (Baltay et
al. 2007).  Because the latter is based on measurements for $\lambda
\le 6000 ${\AA } and only estimated at longer wavelengths, the
$S(\lambda)$ in Figure ~\ref{fig-filter} should be considered
approximate.  The sharp decline at $\lambda \sim7000${\AA } has the
desirable property of eliminating the fringing that plagues
observations that are obtained at longer wavelengths.  Defining the
effective wavelength as $\lambda_{eff}= \int_0^\infty \lambda
S(\lambda)\,\mathrm{d}\lambda$, we obtain $\lambda_{eff} = 5690${\AA }
and $2880${\AA } for the full-width at half maximum (FWHM).  Also
plotted in Figure~\ref{fig-filter} is the $S(\lambda)$ of the Johnson
V pass-band \citep{bessell90}, from which we obtain, for comparison,
$\lambda_{eff} = 5510${\AA } and FWHM $= 850${\AA }.  Although the
$S(\lambda)$ of the LSQ survey is offset in $\lambda_{eff}$ and is
considerably broader than that of the standard V pass-band, the
instrumental magnitudes can be accurately transformed to V magnitudes
using a simple color term.  This is possible because the RRLS span a
fairly narrow range in color, even as they vary in color during their
pulsations.

The raw data of the variability survey are automatically processed by
the Yale Photometric Pipeline, as soon as they start arriving every
morning. This pipeline produces a catalog of positions and
instrumental magnitudes for all objects above a detection threshold.
The software starts with the standard tasks of using the nightly flats
and darks to do background subtraction and flatfielding of the
image. The preprocessed data are then run through SExtractor, a
standard Source Extraction program \citep{bertin96}. In the RR Lyrae
study we used aperture photometry with a 6 pixel diameter
apertures. Instrumental magnitudes were calculated as

    m = $25.0-2.5$log(flux in aperture)

where the flux in the aperture was in units of analog to digital
converter units (ADU). Astrometric calibration was done using the
USNO-B1.0 catalog as the reference.  The precision of the resulting
astrometry was a few tenths of an arcsec in the star positions.

\section{Selection and Calibration of the RR Lyrae Sample}
\subsection{Selection of RR Lyrae Candidates}

Previous RRLS surveys of the halo \citep{kinman66,vivas04} have shown
that RRLS are relatively rare objects on the sky, typically few per
square degree.  This is a consequence of the low density of the halo,
the relatively short time-scale of the RRLS phase, and the fact that the
evolutionary paths of some horizontal branch (HB) stars never enter
the instability strip.  RRLS can be separated from most other kinds of
stars by their variability, but to make this manageable, it is very
useful to limit the possible candidates on the basis of color.  Here
we make use of u-g , g-r colors from the Sloan Digital Sky Survey
(SDSS), as have, for example, \citet{ivezic05, watkins09, sesar10}.

\begin{figure}
\epsscale{0.85}
\plotone{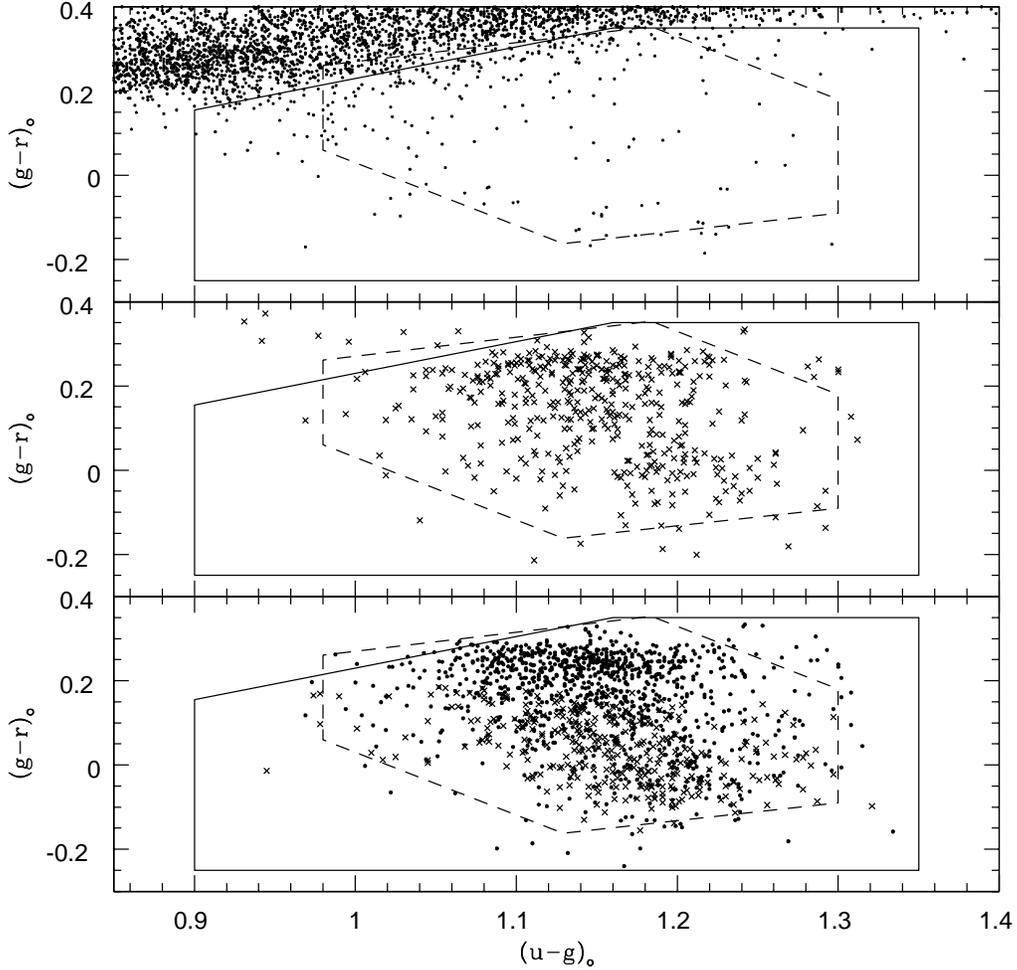}
\caption{The locations of stars in the $(u-g)_0 ,(g-r)_0$ plane and the
  selection of RRLS candidates.  The solid lines enclose the area in
  color space that we used to identify RRLS candidates.  The dashed
  lines enclose the area that \citep{ivezic05} used.  The top diagram
  plots the stars in a small area ($9\ \mathrm{deg}^2$) of the LSQ RRLS survey.
  The large number of stars across the top of this diagram are
  main-sequence stars.  In the middle diagram, the RRLS identified by
  the QUEST survey \citep{vivas04} are plotted.  In the bottom
  diagram, the type ab (solid circles) and type c (x's) RRLS that are
  identified in this investigation (Table 1) are plotted.}
\label{fig-colorbox}
\end{figure}

The top diagram in Figure~\ref{fig-colorbox} shows for a small
fraction of our survey area the locations of stars in the plane defined
by the dereddened colors $(u-g)_0 $ and $(g-r)_0$.  The values
of the interstellar extinctions that the SDSS lists for
each star were used to remove the reddenings.  The large number of
stars lying across the top of the diagram are part of the locus of
main-sequence stars \citep[see][for a larger view]{ivezic05}.  The
stars scattering below at smaller values of $(g-r)_0$ include RRLS.
This is illustrated in the middle diagram in
Figure~\ref{fig-colorbox}, where we have plotted the colors of 441
RRLS from the QUEST survey \citep{vivas04} that we identified in data
release 7 (DR7) of the SDSS survey.  The hexagon that is plotted in
all three diagrams of Figure~\ref{fig-colorbox} encloses the area that
\citet{ivezic05} recommended for finding RRLS candidates based on a
smaller number of QUEST RRLS.  We found that a bigger region (see
Figure~\ref{fig-colorbox}) yields a more complete sample of RRLS.  In
addition, one corner of the hexagon extends into the region occupied
by many main-sequence stars.  While there is indeed some overlap
between RRLS and the main sequence in this plane, we found that a
color selection that is based on the hexagon returns a huge number of
main-sequence stars, a significant fraction of which are variable but
not RRLS.  With little loss of completeness, we found that cutting off
the corner of the hexagon greatly speeded up our survey.  The search
box that we adopted is shown as solid lines in
Figure~\ref{fig-colorbox}, which encloses 98\% of the RRLS from the
QUEST survey.  The colors of the stars in our catalogue of RRLS are
plotted in the lower diagram of Figure~\ref{fig-colorbox}, where one
can see that a few stars lie outside our search box, but inside the
hexagon.  This is a consequence of starting the survey with the hexgon
and later switching to the larger region and repeating the areas
previously covered.

With our 60s exposures, stars brighter than $r= 14$ saturate most of the
CCDs in the camera, and stars fainter than r $\sim 20$ produce low
signal to noise images.  Consequently, we imposed r magnitude limits
of 14 and 20 when selecting candidate RRLS from the SDSS photometric
catalog.  A query of the SDSS database for stars that have the proper
colors to the RRLS candidates yielded more than 29,000 objects in the
region of the sky covered by this first band of the LSQ survey.  The
LSQ database was then searched for these stars, and for each one
found, a separate catalogue of observations was compiled.  After
experimenting with known RRLS, we found that 12 or more observations
were required before a RRLS could be identified with some confidence.
The initial set of candidates was then reduced to the $\sim24,000$
that have at least 12 observations.

\begin{figure}
\epsscale{1.0}
\plotone{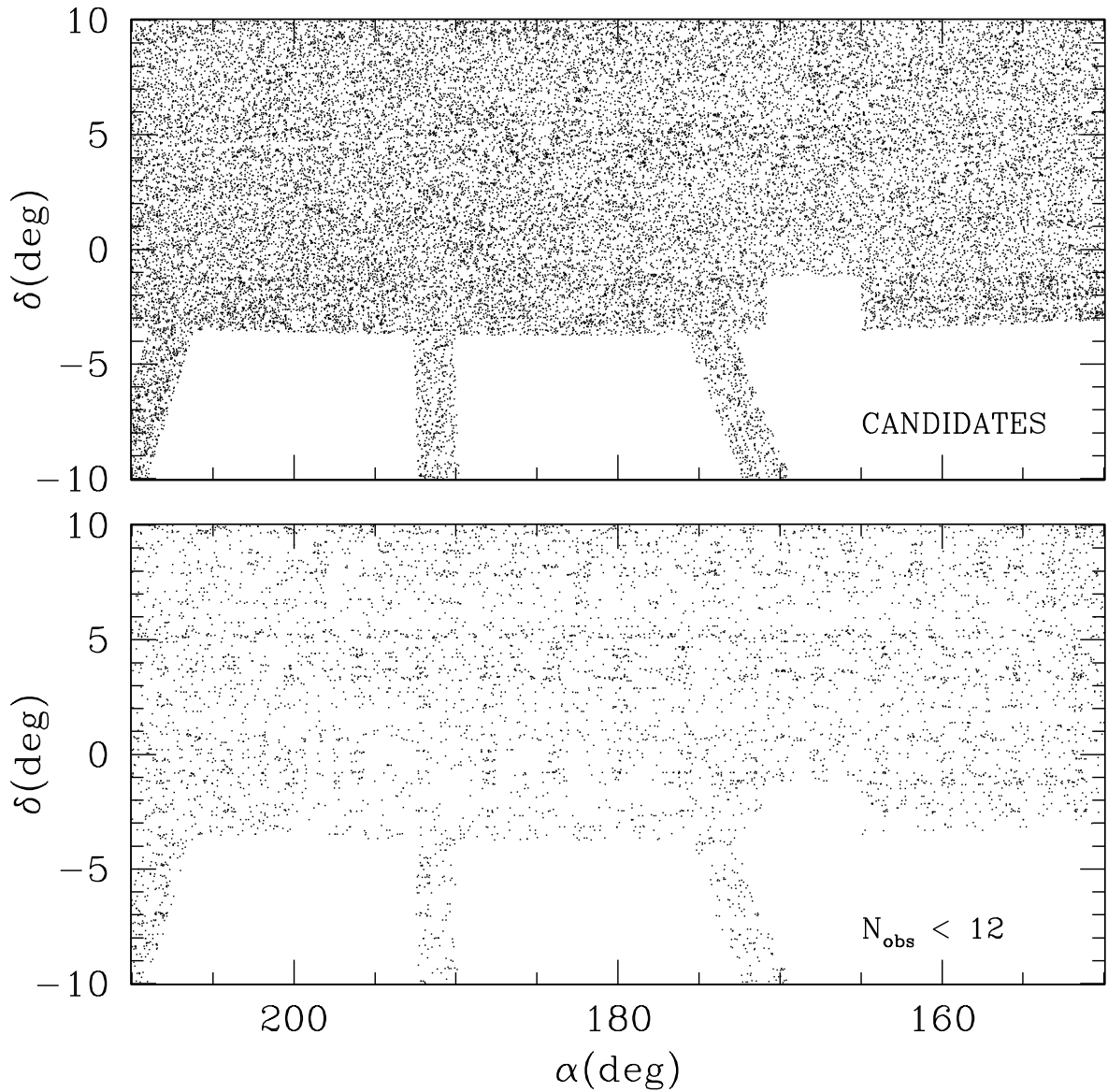}
\caption{The ranges $\alpha$ and $\delta$ included in our RRLS survey.
  In the top diagram, all of the candidate RRLS are plotted.  The
  shape of the survey region was dictated by the footprint of the SDSS
  in this area of the sky.  In the bottom diagram, the candidate RRLS
  for which fewer than 12 observations were obtained are plotted.
  These stars, which for the purpose of identifying RRLS were
  ``unobserved'', constitue 20\% of all candidates.}
\label{fig-skycoverage}
\end{figure}

The area of the sky that is covered by this first part of the LSQ
survey is shown in Figure~\ref{fig-skycoverage}.  Since we select our
candidate RRLS on the basis of their $(u-g)_0$ and $(g-r)_0$ colors,
the survey area is the SDSS footprint between 10 and 14 hrs
($150.0\degr-210.0\degr $)in right ascension ($\alpha $) and
$-10\degr$ and $+10^0$ in declination ($\delta $).  In galactic
coordinates (l,b), the survey has a fan-like geometry that spans
$230\degr<l<350\degr$.  At $l\sim240\degr$, the range in b is $\sim
40\degr - 60\degr$.  Near the middle of region ($l \sim 290\degr$),
the range is $\sim 58\degr - \sim 72\degr$, and at $l\sim 330\degr$ it
is $\sim 57\degr - \sim 70\degr$. The survey region, which covers
$\sim 840 \textrm{deg}^2$ of the sky, contains parts of the leading
stellar stream from the Sgr dSph galaxy, the Virgo Overdensity (VOD),
the Virgo Stellar Stream (VSS), the stellar stream detected by
\citep{walsh09,jerjen13}, and several irregularies in the density
distributions of halo tracers
\citep[see][]{keller08,keller09,keller10} that may be additional halo
substructures.

Plotted in the top diagram of Figure~\ref{fig-skycoverage} is the entire
sample of RRLS candidates that were identified in SDSS DR7.  The gaps
between the rows of CCD's, the loss of several CCD's in the camera,
and the variable weather conditions all conspired to cause some
candidate RRLS to be under observed.  The bottom diagram shows the
locations of the stars for which the LSQ survey obtained fewer than 12
observations.  It shows that these stars, which for our purpose were
essentially unobserved, are scattered throughout the region, with some
concentration in narrow bands in $\delta$.  The stars for which we
did obtain 12 or more observations, the ``observed stars'', constitute
80\% of all the RRLS candidates.

\begin{figure}
\epsscale{1.0}
\plotone{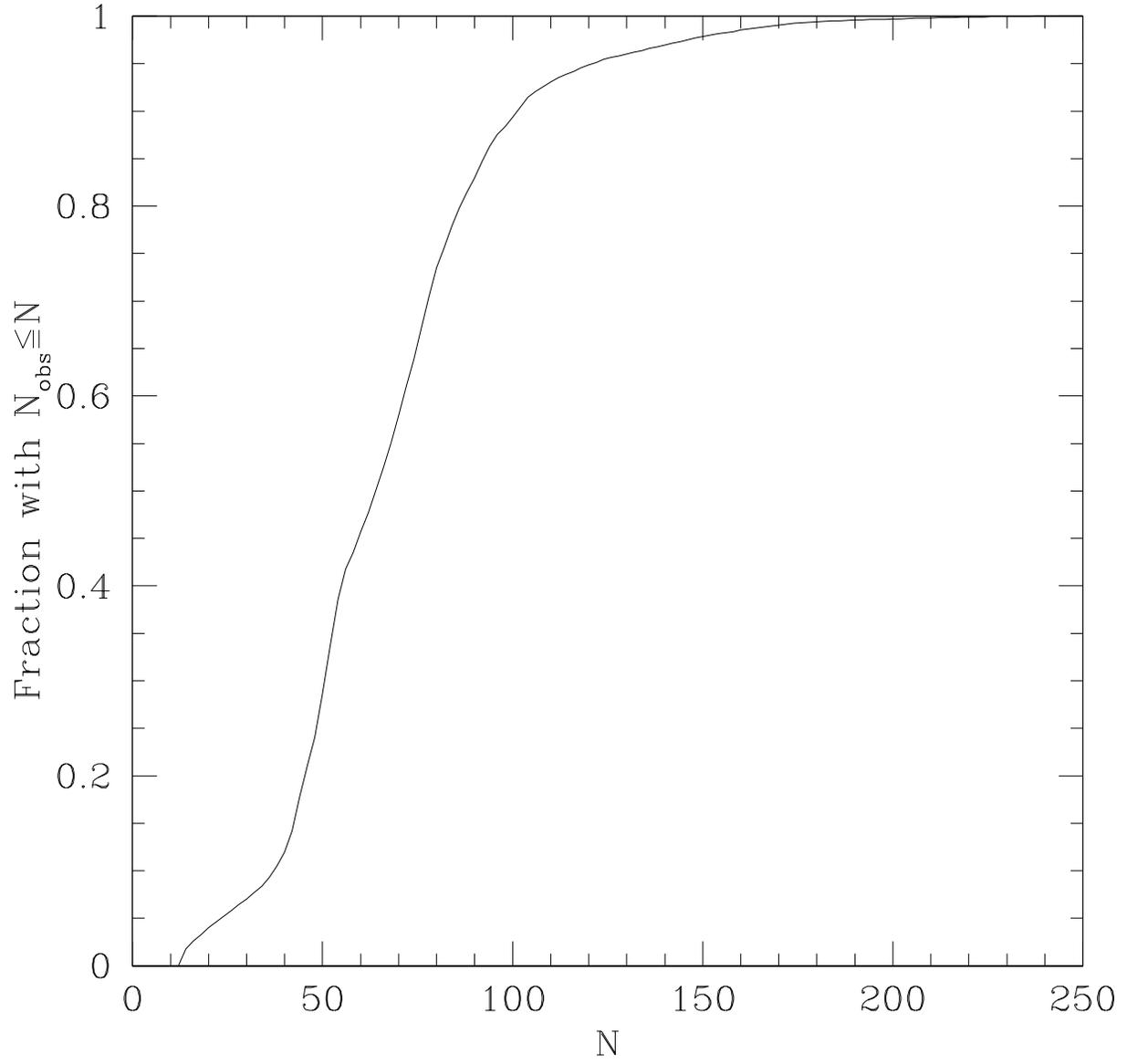}
\caption{The fraction of the sample of ``observed'' candidates (i.e.,
 $ N_{obs} \geq 12$) that have less than N number of observations is
  plotted against N.  More than 50\% of the sample has $N_{obs} \geq 60$.}
\label{fig-cumlative}
\end{figure}

For these observed stars, we plot in Figure~\ref{fig-cumlative} the
fraction of the sample having numbers of observations ($N_{obs}$) less
than a given number N.  One can see from this diagram that 90\% of the
sample of stars have $>35$ observations and 50\% have $>60$
observations.  We discuss below comparisons between the LSQ survey and
previous ones in the same region, which suggest that the very few RRLS
are missed by the LSQ survey if $N_{obs} > 30$.

\subsection{Relative Photometry}

Before identifying the RRLS in the sample of candidates, we performed
relative photometry in order to remove the variations in the
instrumental magnitudes that are produced by the different
sensitivities of the CCD's and the variable atmospheric extinction and
transparency.  For each observation of a candidate RRLS, we saved in a
``standard star'' file the photometry of all point-sources on the same
chip that were more than 6 pixels from the edge of the CCD and had
instrumental magnitudes in the range 12--18, which corresponds to V
$\sim14.2 - 20.2$.  This standard star file was pruned of objects that
contained any bad pixels within a 6x6 pixel box centered on the
object.  One observation was selected as the ``reference observation''
to which all others would be adjusted by differential photometry of
the standard stars.  To select the reference observation, we first
identified the CCD on which there were the most observations of the
target with very reliable astrometry, defined as having a match of
more than 50 stars with the USNO-B1.0 catalog.  Among these
observations from the same CCD, we selected as the reference
observation the one that had the brightest instrumental magnitudes of
the USNO stars, which is likely to be the one that was obtained with
the smallest atmospheric extinction and the highest transparency.
Because the observations from different CCD's could not be as
precisely calibrated as the ones from the same CCD, this procedure
maximized the number of well-calibrated observations.  To determine
the offset between an observation and the reference one, we selected
the standard stars that lie within 6 arcmin of the candidate RRLS.
The median of the differences in magnitude of these standard stars
between the observation and the reference observation was used as the
first estimate of the magnitude offset.  To remove distant outliers,
which may be variable stars, the stars that were more than 6 times the
errors in their magnitudes from the median were clipped from the
sample.  As the final offset, we adopted the mean of the magnitude
differences, weighted by the errors in the magnitudes, of the stars
that remained after the clipping.  This procedure yielded magnitude
offsets with precisions of typically 0.008 mag, which when applied to
the observations of the candidate RRLS, removed their zero-point
differences.  A file was created for each candidate RRLS, which
contained the modified Julian Date of the observations, the zero-point
adjusted instrumental magnitudes and their errors, flags produced by
SExtractor that indicated a potential problem with the measurements,
and a flag indicating the star was within 6 pixels of the CCD edge.
While not all of the flagged observations turned out to deviate
substantially from non-flagged ones, the flags were useful for
identifying the origin of a few very deviant points from the mean
magnitude of a constant star or from an otherwise well-determined
light-curve of a variable star.

\subsection{Identification of RRLS} 

\begin{figure}
\epsscale{1.0}
\plotone{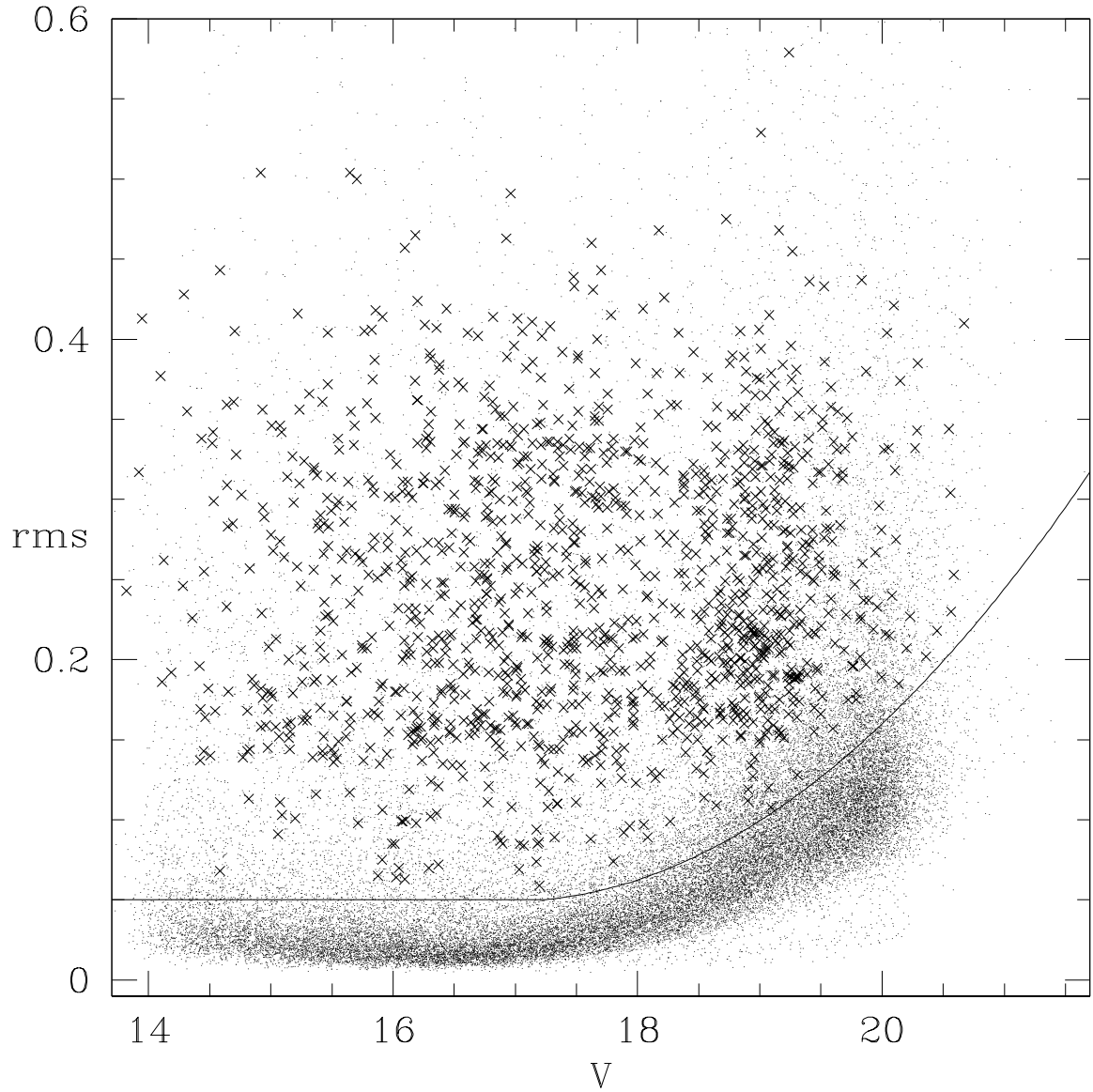}
\caption{The variation of rms with $\langle V \rangle $ for the
  ``observed'' candidates (dots). The solid line is the cut made on
  rms to separate variable from non-variable stars.  The x's are the
  stars identified as RRLS or anomalous Cepheids (Tables 1 \& 2).}
\label{fig-rms}
\end{figure}

Our methods for separating RRLS from non-variable stars and other types of
variable stars closely follow the techniques that we developed for the
QUEST RRLS survey \citep{vivas04}.  For each candidate RRLS found in
the LSQ database, the mean instrumental magnitude and root-mean-square
(rms) of the variations around this mean were calculated.
Figure~\ref{fig-rms} shows a plot of the rms values against a rough V
magnitude, which was obtained by addition of the average zero-point offset of
2.2 to the instrumental magnitudes.  The majority of the stars define
a sequence that is roughly constant at rms$\sim0.02$ over the range
$13.9 < V < ~17.2$.  Fainter than 17.2, this sequence broadens and rises
to rms$ \sim0.12$ at V $\sim20.4$.  To remove constant stars and very
low amplitude variables from consideration, we examined for
periodicity only those stars whose rms values lie above the curve in
Figure~\ref{fig-rms}, which is constant at 0.05 for $V < 17.2$ and is
a quadratic function of V for $V > 17.2$.  The RRLS that we have
identified in this survey are also plotted in Figure~\ref{fig-rms} as the large
symbols, and their locations suggest that the lower limit that we have
set on rms has not excluded many RRLS from further analysis.  In fact,
we had originally set the rms limit 0.01 mag larger.  Lowering it to
the level shown in Fig. 6 yielded $\sim1000$ more candidates, but only
3 of these passed our criteria for RRLS.

To determine if a candidate is an RRLS, we used the period-finding
method of \citet{lafler65} \citep[see also][]{vivas04,saha90}.  The
measurements for each candidate were phased with several thousand
trial periods in the range $0.15 \le P \le 0.9$ days.  For each trial
period, the observations were sorted by increasing phase and the
parameter $\Theta$ was calculated.  $\Theta =
\sum_{i}(m_{i}-m_{i+1})^2/\sum_{i}(m_{i}-\bar{m})^2$ , where
$m_{i}$ and $\bar{m}$ are the individual magnitudes and the mean
magnitude, respectively.  If a trial period produces a light-curve
rather than a scatter plot, $\Theta$ will have a relatively small
value.  To identify values of $\Theta$ that are potentially
interesting, the parameter $\Lambda$ was also calculated for each
trial period.  $\Lambda = \bar{\Theta}/{\Theta} $ , where $\Theta$ is
  the value obtained with the trial period and $\bar{\Theta}$ is the
  average value of $\Theta$ obtained with all trial periods.  A large
  value of $\Lambda$ indicates that the trial period produces a
  well-determined light curve.  A lower limit on $\Lambda$ was set
  after experimenting with successively lower values and seeing if any
  potential RRLS were missed.  If the largest value of $\Lambda$ that
  was obtained with the set of trial periods was less than 2.2, the
  star was excluded from further analysis.  If $\Lambda$ was greater
  than 2.2, then the light-curves produced by the 3 trial periods that
  yielded the largest values of $\Lambda$ were examined by eye to see
  if any had the characteristics of either a type ab (fundamental mode
  pulsator) or a type c (first harmonic mode).  In the QUEST survey,
  the threshold on $\Lambda$ was set at 2.5.  We have lowered it here
  because the set of LSQ measurements for some stars included a few
  very deviant magnitudes.  The otherwise reasonable light curve for a
  RRLS star could have an anomalously low value of $\Theta$ because of
  deviant points, which are easily seen if the light curves are
  examined by eye.  By setting the threshold at 2.2, the light curves
  of many more stars that are not RRLS had to be examined, but it is
  less likely that any true RRLS were missed.

The periods producing these 3 best light-curves were almost
always aliases of each other.  In most cases, the light-curve with the
largest value of $\Lambda$ was unquestionably that of a RRLS or another
type of variable (e.g., W UMa eclipsing binary).  In some cases, there
was some ambiguity as to whether the $1^{st}$ or $2^{nd}$ best
light-curve was the correct one.  If it appeared likely that the star is a
RRLS, the well-known correlations among period, amplitude, and
light-curve shape were used to decide between the two periods.
However, in cases where there were fewer than 25 observations, it was
not always clear what was the best choice between the alias periods.
For $< 1\%$ of the RRLS, it was decided that the period that gave the
third largest value of $\Lambda$ was the correct one.

Once a star was identified as a RRLS, the heliocentric Julian Dates of
the observations were calculated, and template light-curves from
\citet{layden98} or from \citet{mateu09}, for the types ab and c
respectively, were fit to the data.  Allowing the period and the time
of maximum light to vary around the values obtained from the best
trial period, the $\chi^2$ of the deviations of the data from the
templates were minimized, which produced the best estimates of the
period, amplitude, and Heliocentric Julian Date of maximum light.  The
magnitudes of each template were translated to intensity units,
integrated, and the mean was translated back to magnitude units, which
yielded the mean instrumental magnitude, $\langle V_{Inst} \rangle $,
which was later transformed to $\langle V \rangle $ by the calibration
that is described below.  We did not attempt to subdivide the type c
variables into the type d, the multimode variables, or type e, the
suspected $2^{nd}$ harmonic pulsators \citep{alcock96}, because this
does not affect their use as standard candles to probe the halo.

For $\sim 20\%$ of the RRLS, the light-curve produced by the best
trial period had a few points that were clearly very deviant from the
majority.  In most cases, these points had significantly larger
photometric errors than the majority, suggesting that they were
measured from observations with high backgrounds from Moon-light or
through thin clouds.  In some cases, the photometric errors of the
deviant points were small compared to their deviations from an
otherwise well determined light-curve.  The origin of these deviations
are unknown.  They could be due to measurements too near an edge of a
CCD or from behind the joint between the two halves of the filter.  In
all cases, these deviant points and the ones with large photometric
errors had little effect on the identification of a RRLS or the
determination of its period.  They were deleted from the sample before
the template fitting process to ensure that they did not affect the
determination of the best period and $\langle V_{Inst} \rangle$.

\subsection{Absolute Calibration}

The next step in the analysis was to transform $< V_{Inst} >$ to
$\langle V \rangle $ for each RRLS.  The observation that was selected
as the reference image for the differential photometry was calibrated
to V using the photometry of the SDSS DR7 catalogue.  Field stars were
selected within a circle of 300 arcsec on the reference image of each
RR Lyrae star.  The SDSS g and r magnitudes of these stars were then
converted to the Johnson V magnitudes by the relation (Lupton 2005) :

       $V = g-0.5784(g-r)-0.0038$

The mean difference between these magnitudes and their LSQ
instrumental magnitudes define a zero-point offset.  To examine the
accuracy of this technique, we used the same method on 104 stars that
are listed by Stetson (2000) as V standards.  These stars, which have
$14.75 < V < 18.75$, were selected on the basis of being far from CCD
saturation, having good S/N, and having no near neighbors.  Given the
broad width of the LSQ filter, we were not surprised to find that
there was a residual offset that depended on the colors of the stars.
Consequently, we adopted the procedure of removing stars that were
very different in color from RRLS from the samples of calibrating SDSS
stars, and then added the following small correction to the V
mags obtained from the Lupton equation before computing the
zero-point offset:

        $\Delta V=0.1213(g-r)-0.0681$

\begin{figure}
\epsscale{1.0}
\plotone{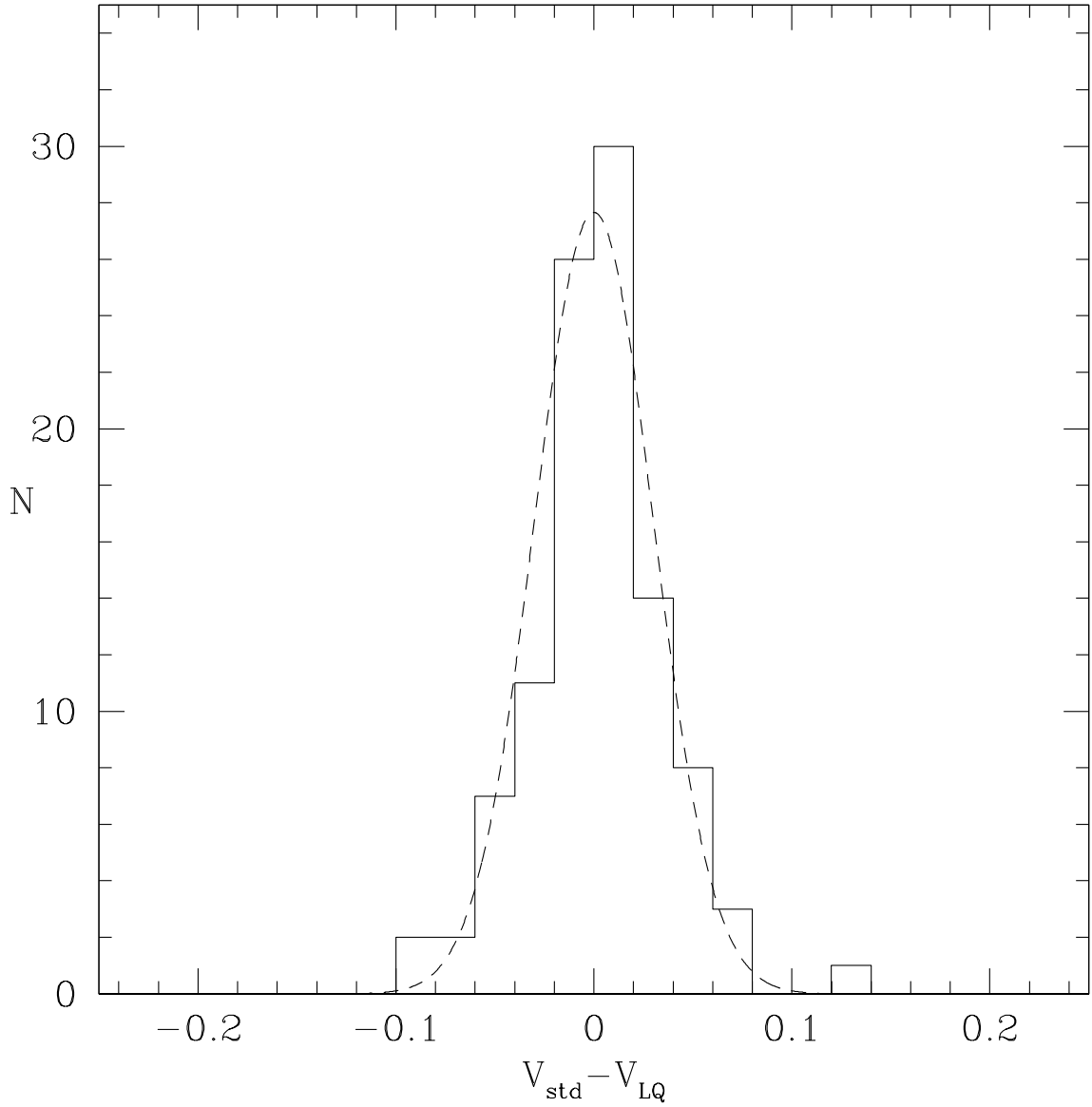}
\caption{For 104 standard stars \citep{stetson00}, the histogram of
  the differences between the standard V and the magnitude obtained
  here.  The dashed curve is a Gaussian distribution with a mean of 0
  and $\sigma = 0.03$.}
\label{fig-stds}
\end{figure}

The success of this technique is evident in Figure~\ref{fig-stds},
where we plot for the Stetson standards the difference between the
Stetson V mag and the V mag obtained from our calibration of the LSQ
observations.  This distribution is well fit with a Gaussian
distribution with a mean of zero and a sigma of 0.03 mag.  Another
estimate of the precisions of our measurements of $\langle V \rangle $
for RRLS is provided by the comparisons between our survey and
previous ones for the same stars (see below).

The filter that is used in the LSQ survey is physically two identical
filters that butt together.  We could not calibrate well the
observations obtained with the CCDs that lie behind the filter joint,
although we could measure the periods and the light-curves of the RRLS
observed with them.  Because of their uncertain calibration, these
stars are labeled with flag = 1 in Table 1.
A comparison with the QUEST survey (see below) suggests that the
calibration for these stars has a sigma of $\sim 0.2$ mag.  The
well-calibrated stars have flag = 0 in Table 1.

Table 1 lists the RRLS by increasing $\alpha$.  The missing numbers
are stars that we believe are Anomalous Cepheids in the Sextans dSph
galaxy (see below), which are listed in Table 2.  In the ``Notes''
column of Table 1, the RRLS that are likely members of the Sex dSph
galaxy are identified, and the variable number is given if the star is
in the study by \citet{mateo95}.  For each star in Tables 1 and 2,
Table 3 lists the heliocentric Julian Date of observation, the observed V
magnitude transformed to the standard system, and its standard
deviation as given by photon statistics.  While the uncertainty in
the calibration to the standard system is larger, in general, that these
standard deviations, they provided a means for weighting the
observations when fitting the template light-curves.

\begin{figure}
\epsscale{1.0}
\plotone{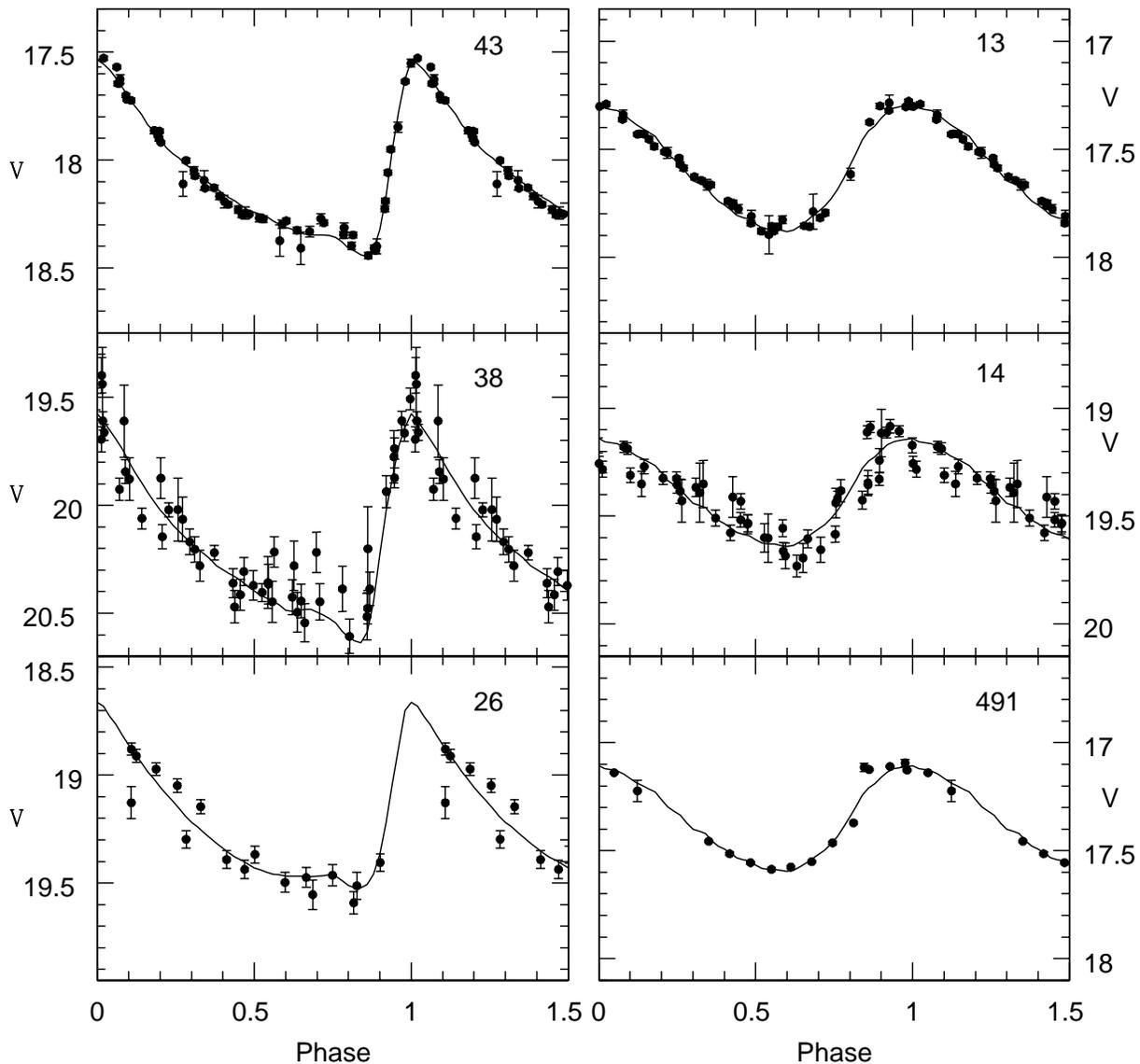}
\caption{Some typical light curves for type ab (left panels) and type
  c (right panels) produced by the LSQ data.  The top panels show
  stars of roughly average brightness and average number of
  observations.  The middle panels show the light curves of stars near
  the faint limit of the survey.  The bottom panels show stars that
  have relatively few observations.}
\label{fig-lcRR}
\end{figure}

In Figure~\ref{fig-lcRR}, a few typical light curves for type ab and
type c variables are displayed on the left and right panels,
respectively.  On the top row are two stars that have mean magnitudes
near the middle of the range of the LSQ survey and typical numbers of
observations ($\sim50$).  The middle row shows two stars near the
faint limit of the survey, again having typical numbers of
observations.  Plotted in the bottom row are two stars that have very
few observations ($< 18$), and yet their light curves are reasonably
well determined.

\subsection{Comparisons with previous RRLS surveys}

In this section, we compare our survey with six previous RRLS surveys
that overlap to some degree.  The comparisons made here are
important check on our detection procedures and our photometric
precision.

\subsubsection{The QUEST Survey}

The LSQ survey includes the region within $150\degr \ge \alpha \le
210\degr$ that was covered by the QUEST survey, which employed an
earlier QUEST camera on the 1 m Schimdt telescope of the Llano del
Halo Observatory in Venezuela.  The original survey covered $380$
$\textrm{deg}^2$ of the sky in a $2.3\degr$ wide band centered on
$\delta = -1$ \citep{vivas04} and $\alpha$ in the ranges $60\degr -
90\degr$ and $120\degr-255\degr$ .  Later a second band at $\delta =
-3\degr$ and $\alpha$ in the ranges $0\degr - 90\degr$ and
$120\degr-195\degr$ was added using the same procedures
\citep{vivas06b}.  Some of the RRLS in the original catalogue have
been re-observed to improve their light-curves (Vivas et al. in
preparation).  The QUEST survey used a V filter, and its measurements
were tied to the standard V system.

In the region of overlap, 223 were identified as RRLS in either the
first or 2nd bands of the QUEST survey.  A few other QUEST RRLS in the
same area of the sky are not included in this comparison because they
are either brighter than the saturation limit of the LSQ survey or are
probably not RRLS on the basis of their SDSS colors and/or a
reexamination of their light-curves.  Of the 223 QUEST RRLS, 178 or
80\% were rediscovered as RRLS by the LSQ survey.  Nineteen of the 45
stars that were missed had fewer than 12 LSQ observations, and all but
2 had fewer than 30.  The QUEST survey is itself incomplete, however
\citep{vivas04}.  In the region of the sky where the LSQ and QUEST
surveys overlap, the LSQ survey identified 283 RRLS, of which 178 are
in common with the QUEST survey.  Thus, the minimum number of RRLS in
this region is 328, and the LSQ and QUEST surveys found 86\% and 68\%
of these RRLS, respectively.  Many of the RRLS that the QUEST survey
missed are near its faint limit, which is consistent with its
estimated completeness \citep{vivas04}.

\begin{figure}
\epsscale{0.8}
\plotone{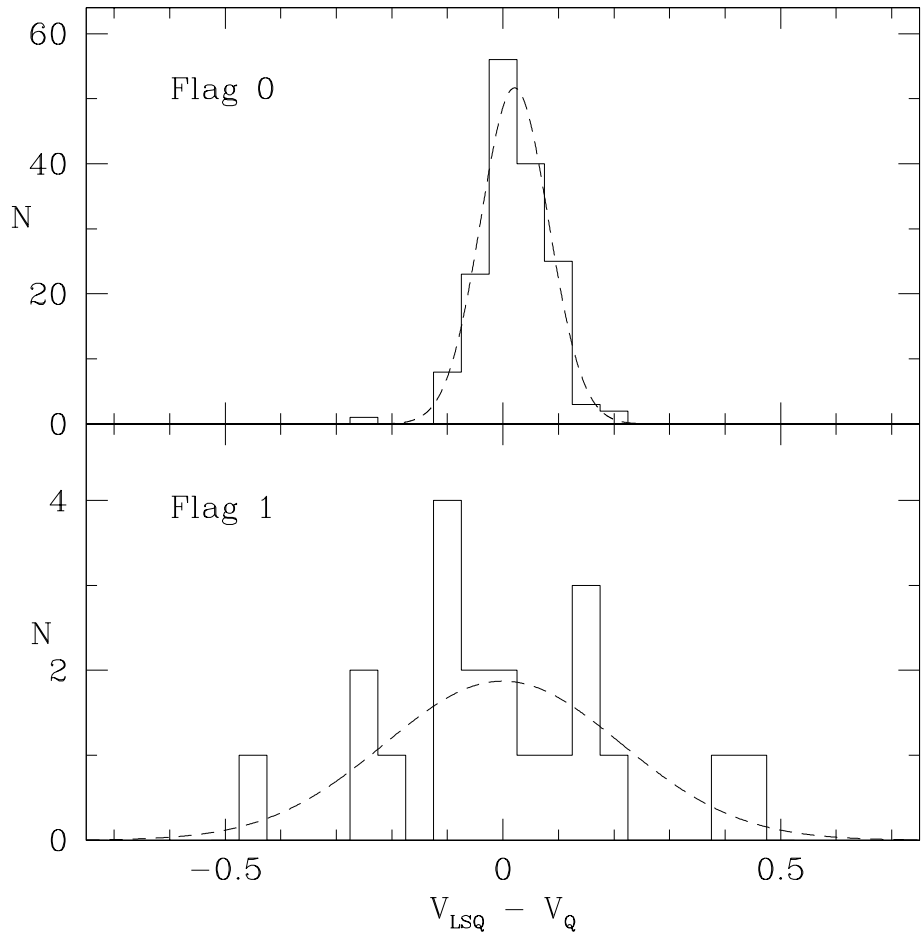}
\caption{The histograms of the differences in $\langle V \rangle $
  between the QUEST survey \citep{vivas04} and our survey.  The top
  diagram compares the 158 RRLS in the QUEST survey for which we
  obtained a good calibration (Flag = 0) of the LSQ data.  The histogram
  is compared with a Gaussian distribution with the same mean (0.021)
  and $\sigma$ (0.061) as the data.  In the bottom diagram, a similar
  comparison is made for 20 QUEST RRLS for which LSQ calibration is
  poor (Flag = 1).  The Gaussian has a mean of 0.006 and $\sigma =
  0.21$.}
\label{fig-LQQuest}
\end{figure}

Of the 178 LSQ RRLS in common with QUEST, 158 have good quality
zero-point calibrations (Flag = 0 in Table 1).  For these stars, the
mean difference in $\langle V \rangle $, in the sense LSQ-QUEST, is
0.021 with a standard deviation ($\sigma$) of 0.061.  The upper
diagram of Figure~\ref{fig-LQQuest} contains the histogram of the
differences, which is compared to a Gaussian distribution that has the
same mean and $\sigma $.  The zero-point uncertainty of the QUEST magnitude
system is 0.02, and in the QUEST catalog the errors in $\langle V
\rangle $ stemming from the light-curve fitting average about 0.05 mag
(Vivas \& Zinn 2006).  Because additional observations have been added
more recently for several stars and the $\langle V \rangle $ values
have been updated, the average error of the QUEST magnitudes may be
0.04.  With this estimate, the agreement between the QUEST and the
well-calibrated LSQ measurements of $\langle V \rangle $ suggest that
the errors in the latter are also $\sim0.04$ mag.

For 20 of the stars in common with QUEST, only a very rough
calibration could be made of the LSQ magnitudes (flag = 1 in Table 1).
The lower diagram in Figure~\ref{fig-LQQuest} shows the distribution
of the differences in $\langle V \rangle $ between LSQ and QUEST,
which is compared with a Gaussian distribution that has the same mean
(0.006) and $\sigma$ (0.21).  The poorly calibrated LSQ measurements
have as expected large uncertainties in their mean magnitudes, and
this compromises to some extent their use as standard candles.  The
identification of these stars as RRLS, and their periods, times of
maximum light, and amplitudes are not sensitive to this uncertainty.

A comparison of the LSQ periods and the QUEST periods indicated very
little difference for most stars ($|\Delta P|< 1x10^{-3}$ day).  However, 5
stars that are identified here as long-period, low-amplitude type ab
(more precisely type b), QUEST had identified as type c.  In each
case, the two periods are aliases of each other, and the one selected
here gives better results than the QUEST period with the LSQ data and
also excellent light-curves with the QUEST data.  According to the LSQ
observations, another QUEST star (Q119) is an Anomalous Cepheid in the
Sextans dSph galaxy, with a period of about 2 days (see below).  The
period found here also fits the QUEST observations.

\subsubsection{The sample of \citet{wu05}}
 
\citet{wu05} obtained unfiltered CCD
 photometry for 69 stars that were identified as candidate RRLS in the
 commissioning data of the SDSS by \citet{ivezic00}.  They
 confirmed that these stars are RRLS and measured their periods and
 light-curves.

In the region of the sky covered by the LSQ survey, \citet{wu05}
measured the light curves and periods of 42 stars.  Thirty-two of
these stars were identified as RRLS by the LSQ survey, and one was
rejected as a probable W UMa eclipsing variable (\#29 in Wu et
al. 2005, which is 211 in the QUEST survey, Vivas et al. 2004). The
recovery rate of RRLS is therefore 78\%.  All of the missed RRLS had
fewer than 20 LSQ observations, and 7 of the 9 had fewer than 10
observations.  The periods derived here are in excellent agreement
with the ones obtained by Wu et al. ($|\Delta P|<1x10^{-3} $day).
Their unfiltered measurements were calibrated to r' band with a
precision of ~0.07 mag, and the transformation between $ \langle r'
\rangle $ and $\langle V \rangle $ was uncertain by ~0.1 mag. because
measurements of mean color were not obtained \citep{wu05}.
Consequently, we have not made a comparison between the mean
magnitudes because it would carry much less weight than the ones
involving the QUEST survey and the ones below.

\subsubsection{The LONEOS-I RRLS survey}

\citet{miceli08} used the data from the Lowell Observatory Near Earth
Objects Survey Phase I (LONEOS-I) to identify 838 ab type RRLS in a
1430 $\textrm{deg}^2$ area of the sky, which partially overlaps with
the LSQ survey.  Their photometry is tied to a synthetic SDSS-based V
magnitude with a zero-point scatter of 0.13 mag.  There are 16
LONEOS-I RRLS in the region of overlap and fainter than the bright
magnitude limit of the LSQ survey.  The LSQ survey recovered 14 or
~88\% of these stars.  For the two stars that were missed, the LSQ
survey did not obtain observations at their positions, probably
because the stars either fell on the gap between two CCDs or on a dead
CCD.  For 13 of the 14 stars in common, the LSQ magnitudes are well
calibrated.  The mean difference in $\langle V \rangle $, in the sense
LSQ-LONEOS-I, is 0.024 with $\sigma = 0.139$.  These differences are
consistent with the estimate of the errors in the LSQ magnitudes that
we made above from the comparison with QUEST, and the estimate made by
Miceli et al. (2008) of the zero-point uncertainty in the LONEOS-I
measurements.  The periods obtained for 14 stars in common are in good
agreement ($|\Delta P|<2\textrm{x}10{-3}$).

\subsubsection{The SEKBO survey} 

Using the data from the Southern Edgeworth-Kuiper Belt Object (SEKBO)
survey, Keller et al. (2008) selected a large number of candidate RRLS
based on color and variability in typically 3 observations.  Later,
Prior et al. (2009) obtained sufficient observations to construct
light-curves for a subsample of these stars, and 7 lie in the region
covered by the LSQ survey.  Only 3 (43\%) were recovered by the LSQ
survey.  Fewer than 10 observations were obtained for three of the
missed stars.  Although 24 observations were obtained of the remaining
star, which is usually sufficient for a good quality light curve, the
LSQ observations are noisy, and a firm identification as a RRLS could
not be made.  For one of the RRLS in Prior et al. (2009; 107434-394 =
LSQ152), their period (0.327 day) from only 8 observations appears to
be incorrect.  The 77 LSQ observations indicate that this star is a
type ab with a period of 0.65578 day.  The periods of the other two
stars in common are in good agreement, and for these two stars the
mean difference in $\langle V \rangle $, in the sense LSQ-Prior et
al., is -0.01 with $\sigma = 0.07$, which are consistent with errors
in the measurements.

\subsubsection{RRLS and Anomalous Cepheids in the Sextans dSph galaxy}

 The Sextans dSph galaxy lies near the southern boundary of our RRLS
 survey at $\alpha = 153.260$ and $\delta = -1.620$.  Its distance,
 $86\pm4$ kpc \citep{mateo98}, places it at the extreme limit of the
 LSQ survey. The positions of the stars that we identify here with
 Sextans are within the tidal radius of its center, according to the
 measurements by \citet{irwin95}.  Since halo RRLS of their $<V>$ are
 very rare, this is strong evidence for membership in the dSph
 galaxy. Our requirement that the candidate RRLS from the
 SDSS have $r <20.0$ undoubtedly removed some Sextans RRLS from
 possible inclusion in our sample.  While we intend later to make a
 more thorough survey of Sextans without this limitation, we note for
 now that our sample includes 7 RRLS that are probably members (Table 1).  It
 also includes 6 stars that are known or likely Anomalous Cepheids
 (AC).  The AC are listed in Table 2 and discussed briefly below.

One RRLS and 3 AC in our sample of Sextans variables were observed
previously by \citet{mateo95}.  The mean difference in $\langle V
\rangle $, in the sense LSQ-Mateo et al., is 0.00, with $\sigma =
0.05$.  Since these stars are among the very faintest in our sample,
it is encouraging that the agreement is this good.  For one of the AC
variables (LSQ34 = V34, Mateo et al.), our observations are better fit
with a period of 0.50640 day than with the period of 0.34126 that is
listed by Mateo et al. (1995).  For the other three stars in common,
the periods are in agreement ($|\Delta P| < 2\textrm{x}10^{-3}$).  The
light-curve of RRLS LSQ38 \citep[V3 in][]{mateo95} is shown in
Figure~\ref{fig-lcRR} .

In Table 2, we have listed the 6 probable AC variables, 3 of which are
in common with \citet{mateo95}.  Their absolute V magnitudes, $M_V$,
were calculated using an apparent distance modulus of $(m-M)_V = 19.76$
for Sextans (Mateo 1998).  According to the relationships given in
Pritzl et al. (2002), these stars lie on either the P-L relationship
for fundamental or first harmonic AC variables.  LSQ 40 is
noteworthy because its $\sim2$ day period is one of the longest periods of
the AC variables in the dSph galaxies.

\subsubsection{The Catalina Surveys}

The recent large surveys for type ab RRLS using the data from the
Catalina Surveys Data Release 1 became available as our analysis was
nearing completion \citep{drake13a,drake13b}.  These surveys, which
cover a large area of the sky, include the region studied here.  The
Catalina surveys used three telescopes, each equipped with similar CCD
detectors without filters.  Their calibration methods for RRLS yielded
measurements of $\langle V \rangle $ with an accuracy of $\sim 0.05$
mag \citep{drake13a}.  A comparison of the positions of the stars in
the LSQ and Catalina surveys yielded a total of 791 stars in common,
790 RRLS and one AC in the Sextans dSph galaxy (LSQ36 = V1 in the
study by Mateo et al. 1995).

There is generally good agreement between the periods determined by
the LSQ and Catalina surveys.  With the exception of 4 stars, $|\Delta
P|< 7\textrm{x}10^{-3}$ day, with an average of $2.4\textrm{x}10^{-4}$.
For one (LSQ611) of the 4 deviant stars, LSQ survey had obtained only 11
good observations.  This star was correctly identified by the LSQ
survey as a type ab RRLS, but the wrong period was selected from the
two that produced equally good light curves.  The other period agrees
with one obtained by the Catalina survey, and we have adopted it for
the LSQ catalogue (Table 1).  The other 3 deviant stars (LSQ360, 522,
953) are classified as type ab RRLS by the Catalina survey, but the
LSQ survey indicates that they have the periods, amplitudes, and
light-curve shapes of type c variables.  We have retained the LSQ
classification for the LSQ catalogue.

For 730 stars in common, the LSQ measurements have good photometric
calibrations (Flag = 0 in Table 1).  The mean difference in $\langle V
\rangle $ in the sense LSQ-Catalina is 0.07 with a standard deviation
of 0.11.  Some of this discrepancy is probably due to image blending,
which affects differently the two surveys.  For example, 22 of the
stars in common differ in $\langle V \rangle $ by more than 0.3 mag.
For 19 of these stars the deviations are positive, indicating that the
Catalina surveys measured brighter magnitudes.  An inspection of the
SDSS images for these stars suggests that the measurements for 17 of
them may be affected by the presence of nearby stars or galaxies.  One
of the 3 stars with a large negative offsets in magnitude is a
relatively bright star (LSQ1063), which is near the saturation limit of
the LSQ survey.  The LSQ measurements of it are therefore suspect.
There is no ready explanation for the other 2 stars with large
negative deviations or for the 2 stars with large positive ones that
appear to be unblended.  The measurements for some of the stars that
deviate by smaller amounts may also be affected by image blending.

For 61 of the stars in common between the LSQ and Catalina surveys,
only a poor photometric calibration was possible for the LSQ
observations (Flag = 1 in Table 1).  The mean difference in $\langle V
\rangle $, in the same sense as above, for these stars is 0.06 mag
with a standard deviation of 0.22 mag.  An almost identical standard
deviation was found above in the comparison between the LSQ and QUEST
surveys.

To investigate the completeness of the LSQ survey, we restricted the
overlap region with the Catalina surveys to $\delta \geq -1\degr$, where the
rectangular geometry of the LSQ survey simplified the comparison.  The
LSQ survey is seriously incomplete for $\langle V \rangle \leq 14.0$
because of saturation of the CCDs, and both surveys are incomplete
fainter than $\langle V \rangle = 20.0$.  Within $14\leq \langle V
\rangle \leq 20$ and for $10\degr \geq \delta \geq -1\degr$, the
Catalina surveys identified 859 type ab variables and the LSQ survey
found 766 type ab and 267 type c RRLS, of which 615 are in common with
Catalina.  The LSQ survey therefore found 72\% of the Catalina type ab
RRLS.  Of the ones that the LSQ survey missed, only 15\% had the 12 or
more observations that were required were before a period search was
undertaken, and only 7\% had more than 30 observations, which is usually
adequate for a good quality light-curve.  The LSQ data for these
relatively well-observed stars indicated that they were variable
stars, but the data were either too noisy or too poorly spaced in
phase for a firm decision to be made about the type of variable.

\subsection{Completeness of the Survey}

The completeness of the LSQ RRLS survey can be estimated from the
overlap with the previous surveys.  As noted above, for 80\% of the
RRLS candidates sufficient observations were obtained for a period
search (i.e., $N_{obs} \ge 12$), and our color selection box includes
98\% of the QUEST RRLS.  The above comparisons between the LSQ survey
and the previous ones indicate that some additional RRLS were missed,
primarily because of small $N_{obs}$.  For stars with $N_{obs} \ge
12$, the LSQ survey found $\sim 92\%$ of the RRLS that were identified by the
other surveys (exluding the survey of the Sex dSph galaxy, which only
contains stars at the faint limit of the survey). If this is
representative of the whole survey region, the LSQ survey contains $
\sim 70\% $ of the total population of RRLS ($\sim 0.80$x0.98x0.92).

\begin{figure}
\epsscale{0.7}
\plotone{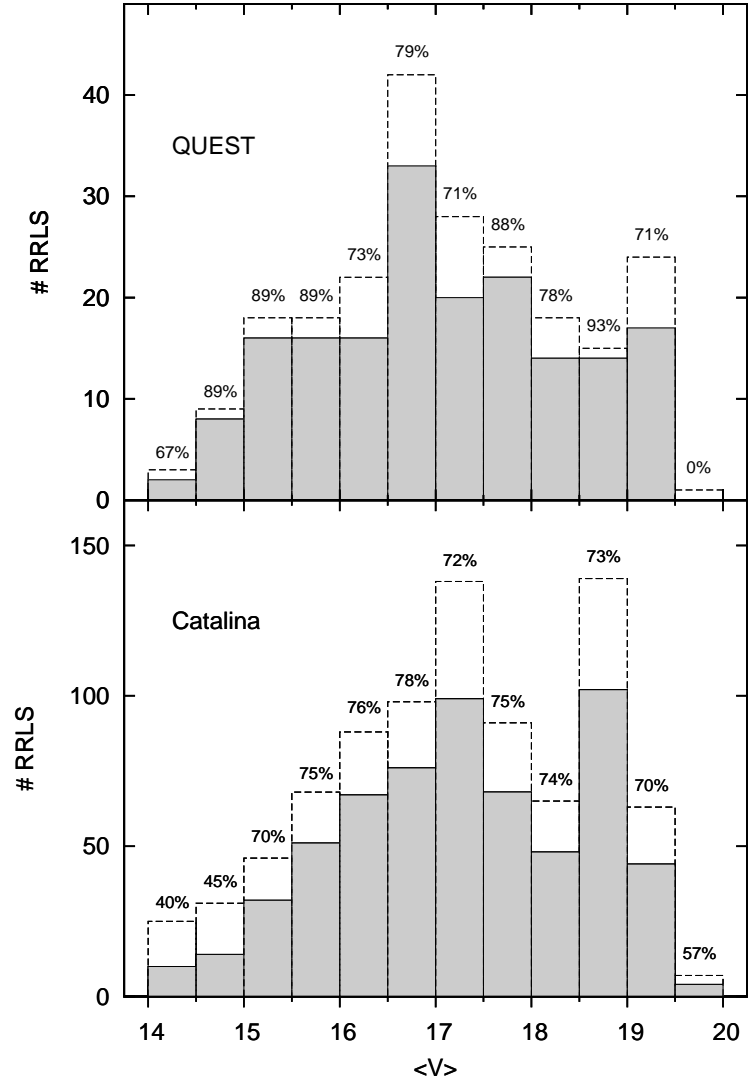}
\caption{In the top and bottom diagrams, the dotted histograms are the
  distributions with $\langle V \rangle $ of the QUEST and Catalina
  RRLS in the regions of overlap with the LSQ survey.  The filled
  histrograms is the distribution of the ones recovered by the LSQ
  survey.  The percentages recovered are shown for each bin.}
\label{fig-Magstats}
\end{figure}

To examine the completeness of the LSQ survey as a function of
$\langle V \rangle $, we show in the top diagram of
Figure~\ref{fig-Magstats} the magnitude distribution of the 223 QUEST
RRLS in the region of overlap and the distribution for the 178 that
were recovered by the LSQ survey.  This diagram illustrates that
$\gtrsim70\%$ of the QUEST RRLS were recovered over the magnitude
range $14 \leq \langle V \rangle \leq 19.5$.  Only one QUEST RRLS in
the overlap region is fainter than 19.5, and it was missed by the LSQ
survey.  In the lower diagram in Figure~\ref{fig-Magstats}, we show a
similar plot for the 859 type ab RRLS identified by the Catalina
surveys in the region of overlap with $\delta \geq -1\degr$ and the
615 RRLS recovered by the LSQ survey in that region.  With the
exceptions of the two brightest bins, this diagram is consistent with
top one in showing that the completion of the LSQ survey is
$\gtrsim70\%$ for $\langle V \rangle \leq 19.5$.

The variables in the Sex dSph galaxy provide additional information on
the completion of the LSQ survey near its faint limit.  Of the 42 RRLS
and AC variables in the study by \citet{mateo95}, 34 are RRLS in the
narrow interval $20.25\le \langle V \rangle \le 20.60$. The LSQ survey
did not recover any of these variables.  The LSQ survey did recover 4
of the 8 remaining variables that span the range $18.88 \le \langle V
\rangle \le 20.25$, and it found 6 new RRLS in the dSph galaxy with
$20.0 \le \langle V \rangle \le 20.5$.  This and
Figure~\ref{fig-Magstats} suggest that the completeness of the LSQ
survey is $\gtrsim70\%$ for $15.0 \lesssim \langle V \rangle \leq 19.5$, but
then falls to zero over the range $19.5 \leq \langle V \rangle \leq
20.6$.  

Because the CCD's in the LSQ camera have different sensitivities and
saturation limits, it is likely that the completeness at $\langle V
\rangle > 19.5$ and $\lesssim 14.5$ varies with position on the sky.
For RRLS with small interstellar extinctions, as is the case in this
region of the LSQ survey, $\langle V \rangle \sim 19.5$ corresponds to
a distance from the Sun, $d_{\sun}$, of $\sim 60$ kpc.  In the
following discussion of the density profile of the halo, we do not use
the densities of RRLS at $d_{\sun} > 60$ kpc in calculations nor do we
compute densities for stars with $\langle V \rangle < 14.55 $.

\section{The spatial distribution of the RR Lyrae variables}

To determine the distances of the RRLS, we have adopted an absolute
magnitude of $M_V = +0.55$ for all stars.  This value, which we used
earlier in the QUEST survey \citep{vivas06a}, is in agreement with most
$M_V - [Fe/H]$ relations for RRLS near the middle of the [Fe/H] range
of halo RRLS, i.e., [Fe/H] $\sim -1.6$, including the recent
determination by statistical parallax for type c variables
\citep{kollmeier12}.  For the interstellar extinction, we adopted the
values given by the dust maps of \citet{schlegel98}.  For the
well-calibrated stars in the LSQ survey (Flag = 0 in Table 1), the
fractional error ($\sigma_{d_{\sun}}/d_{\sun}$) is about 0.07
\citep[see][]{vivas06a}.  For the poorly calibrated stars (Flag = 1), the
fractional error is $\sim 0.11$, which is similar to the distance
uncertainties for halo main-sequence stars \citep[e.g.,][]{bonaca12b}.

\begin{figure}
\epsscale{0.9}
\plotone{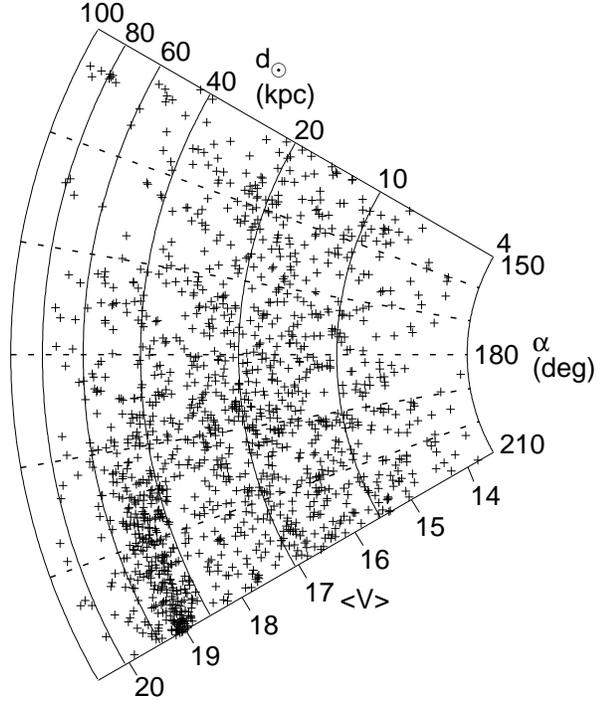}
\caption{For the LSQ RRLS, mean V mag ($\langle V \rangle $),
  corrected for extinction, and the corresponding $d_{\sun}$ is
  plotted against $\alpha$.  The large concentration of RRLS from
  $185\degr \lesssim \alpha \lesssim 210\degr$ with $\langle V \rangle
  \sim 19 $ is the leading stream from the Sgr dSph galaxy.  The small
  concentration of RRLS at $\alpha \sim 153$ and $\langle V \rangle
  \sim 20$ is the Sex dSph galaxy.  The Virgo Stellar Stream (VSS) is
  located near $\alpha \sim 186\degr$ and $d_{\sun} \sim 20$
  \citep{duffau06}, where there is a concentration of RRLS in this
  diagram.}
\label{fig-WedgeData}
\end{figure}

The distribution of the LSQ sample of RRLS as a function of $\alpha$
and $d_{\sun}$ is shown in Figure~\ref{fig-WedgeData}, where one can
readily see the leading stream from the Sgr dSph galaxy and the Sex
dSph galaxy.  While the overdensity due to the Virgo Stellar Stream
(VSS) can be recognized in this diagram, its extent becomes clearer
once the background of RRLS is estimated.  To do this for the LSQ
RRLS, we first calculate the variation of their number density with
$R_{gc}$. Since the form of this density profile provides a clue to
the formation of the halo, it is worthy of investigation on its own.
Recent studies have confirmed older ones that suggested that the halo
is flattened in the same direction as the Galactic disk and have also
greatly strengthened the case made first by the RRLS survey by
\citet{saha85} that there is a significant break in the power-law
slope of the density profile, e.g., \citet{sesar11}, \citet{deason11}.
We find evidence for both of these effects here.

\subsection{The number density profiles}

The sample of RRLS is sufficiently large that it is possible to
subdivide it and still have sufficient numbers of stars in each
subregion to determine the density profile.  We can then investigate
the variation of the density profile with direction.  Because the
geometry of the survey region is simpler in equatorial coordinates
than galactic ones, we divided the survey area into 12 strips, each
$ 5\degr $ wide in $ \alpha $.  These regions vary somewhat in $ \delta $
(see Figure~\ref{fig-skycoverage}), which was taken into account when
calculating the number density ($\rho$ in units of $\# kpc^{-3}$) of
RRLS.  We also tried wider strips in $\alpha$ with subdivisions by
$\delta$ and did not find any significant changes in the profiles.  To
avoid sample incompleteness due to either the saturation of the CCDs
or the bright cutoff in the candidate selection, we excluded the very
brightest RRLS when calculating the number densities.  The number
densities were calculated in bins in distance modulus, $(\langle V
\rangle - M_V)_0$, starting at 14.0.  The width of the first bin was
1.0 mag to ensure that the volume was sufficiently large that some
RRLS were included.  Subsequent bins were 0.5 mag wide.  Since the
first eight bins only include stars with $d_\sun \lesssim 50$ kpc,
they should be immune from the incompleteness near the sample's faint
limit (see above), and only these bins were used when fitting a
function to the number densities.  The plots of the number densities
include more distant bins because it is interesting to see how they
compare with extrapolations of the fitted functions.

One can see from Figure~\ref{fig-WedgeData} that the subdivisions by
$\alpha$ will be affected differently by the major substructures in
the region.  Obviously, the Sgr stream will have a profound effect on
the profiles for $\alpha \gtrsim 185\degr$.  The VSS is expected to
have a smaller but nonetheless significant effect on the profiles for
$180\degr \lesssim \alpha \lesssim 190\degr $.  Because the region
$150\degr \le \alpha \le 170\degr$ appears to be devoid from major
substructures, it was chosen for estimating the density profile of the
smooth component of the halo, but first the RRLS from the Sex dSph
galaxy were excluded.

\begin{figure}
\epsscale{0.7}
\plotone{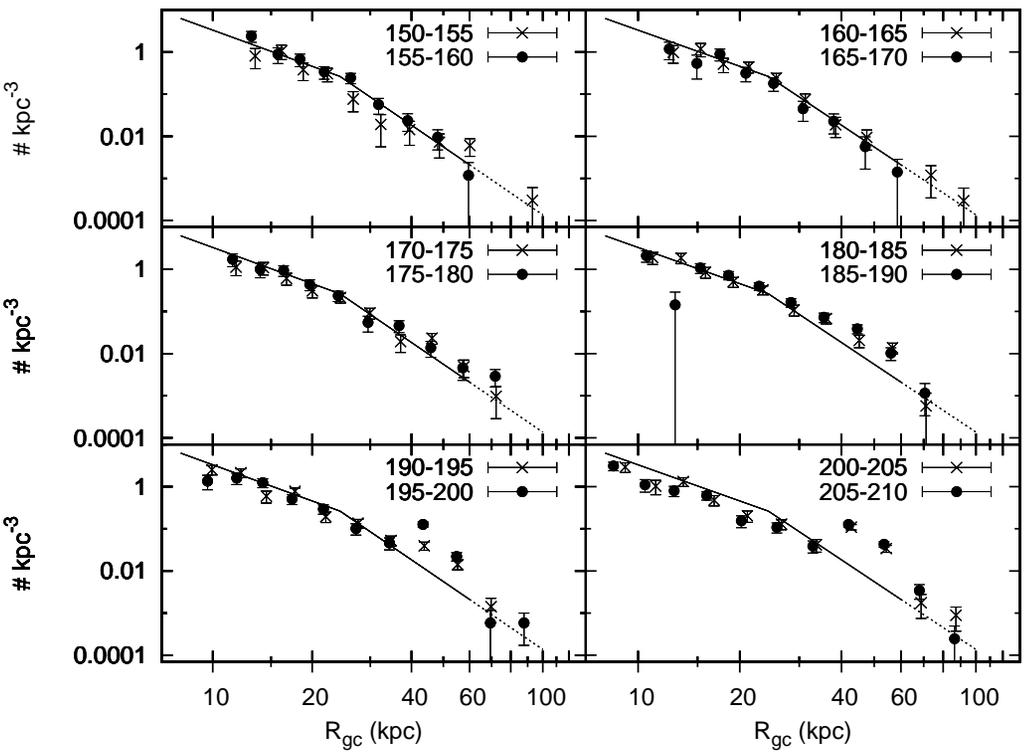}
\caption{The number desity of RRLS is plotted against $R_{gc}$ for
  different ranges in $\alpha$.  The lines are fits to the data over
  the range $150\degr \leq \alpha \leq 180\degr$.  The first 4 radial bins and
  next 4 radial bins have been fit separately because of the break in
  the density profile at $R_{gc} \sim 25$ kpc.  A dashed line
  indicates extrapolation past $d_{\sun} = 60$, where incompleteness
  may be significant.}
\label{fig-denRgc}
\end{figure}

As others have in the past \citep[e.g.,][]{preston91,vivas06a}, we
adopt either a spherical halo or an ellipsoidal one that is flattened
towards the galactic plane.  For the density profile, we adopted
$\rho(R) = \rho_\sun(R/R_\sun)^n$, where R is either the radius of the
sphere or the semimajor axis of the ellipsoid, and $R_\sun$ and
$\rho_\sun$ are the distance of the Sun \citep[set to 8
  kpc, e.g.,][]{majaess10} and the density at the Sun's position,
respectively.  The logarithmic form of this equation (eq. 1) was used
to make weighted least squares fits to the densities in the different
bins of $(\langle V \rangle - M_V)_0$.  For a spherical halo, which we
will consider first, $R = R_{gc} = (X^2 + Y^2 + Z^2)^{1/2}$ , where X,
Y, and Z are cartesian coordinates with the Galactic center at the
origin.  The X and Y axes are in the Galactic plane, and the Z axis is
perpendicular to the plane.  The top diagrams in
Figure~\ref{fig-denRgc} are plots of the densities in the $5\degr$
subdivisions in the range $150\degr \le \alpha \le 170\degr$, and they
show that the density profile has a pronounced break at $R_{gc} \sim
25 \textrm{kpc}$ in each subdivision.  Combining the data for these 4
subdivisions yields 16 density measurements on each side of the break.
Fitting eq 1 separately to these regions, we obtained the following
broken power-law profile for the RRLS, and the reduced chi-square,
$\chi^{2}_{\nu}$.

\begin{eqnarray}
log(\rho) = log(\rho_\sun) + nlog(R/8.0)   \\
R = R_{gc} \nonumber \\
R_{gc} < 24.4 \ \textrm{kpc} \nonumber \\
log(\rho_\sun) = 0.80\pm0.17,  n = -2.8\pm0.5,  \chi^{2}_{\nu} = 0.67 \nonumber  \\
R_{gc} > 24.4 \ \textrm{kpc} \nonumber \\
log(\rho_\sun) = 2.00\pm0.31,  n = -5.4\pm0.5, \chi^{2}_{\nu} = 0.54  \nonumber 
\end{eqnarray}

Top panels in Figure~\ref{fig-denRgc} show that this profile is a good
fit to the number densities even beyond the distance where the
incompleteness of the sample is expected to grow.  To estimate the
density at the Sun, the profile has to be extrapolated to 8 kpc, which
yields $\rho_\sun = 6.2 \pm 3.3 \textrm{kpc}^{-3}$.  If, as suggested
above, the completeness of the LSQ survey is $\sim 0.7$, then the true
density is $\sim 8.9 \textrm{kpc}^{-3}$ at the Sun.  Multiplying this
number by the fraction $\sim 0.74$ of type ab variable in the sample,
we obtain $6.6 \textrm{kpc}^{-3}$ for type ab variables alone, which
is higher than some recent estimates (e.g., $4.2 \textrm{kpc}^{-3}$
Vivas \& Zinn 2006 and $5.6 \textrm{kpc}^{-3}$ Sesar et al. 2010), but
within our errors of them.

The middle left panel of the Figure shows that this profile
approximately fits the data in the range $170\degr \le \alpha \le
180\degr$ before the break, (i.e., $R_{gc} < 24.4$\ kpc) but it
declines somewhat too steeply beyond the break.  The same can be said
for the next subdivision, $180\degr \le \alpha \le 190\degr$, however,
this region includes the VSS and the Sgr stream.  The bottom panels
show that starting in the range $190\degr \le \alpha \le 200\degr$ and
becoming more pronounced in $200\degr \le \alpha \le 210\degr$, the
profile runs roughly parallel to the data before the break, but it is
offset to higher densities.  A systematic deviation from spherical
model is expected in this sense if the halo flattened towards the
plane.  The huge overdensity caused by the Sgr stream in these panels
makes any comparison between the profile and the data essentially
meaningless beyond the break.

The relatively small range in the directions probed by the survey
limit its sensitivity for determining the flattening parameter q, the
ratio of the minor axis (Z), to major axes (X and Y) of an ellipsoidal model.
We can, however, see if a particular halo model fits the data, and we
chose to examine the recent one by \citet{sesar11} in which q = 0.7.
Following their nomenclature, $R_e$ is the semimajor axis of the
ellipsoid, and it is substituted for R in equation 1.  We obtained the
following results fitting, as before, the densities in the region $150\degr
\le \alpha \le 170\degr$.

\begin{eqnarray}
R = R_e = (X^2 + Y^2 + (\frac{Z}{0.7})^2)^{1/2} \nonumber \\
R_e < 29.5 \ \textrm{kpc}  \nonumber \\
log(\rho_\sun) = 0.85\pm0.18 ,n = -2.5\pm0.4, \chi^{2}_{\nu} = 0.69 \nonumber \\
R_e > 29.5 \ \textrm{kpc} \nonumber \\
log(\rho_\sun) = 2.35\pm0.35, n = -5.2\pm 0.5, \chi^{2}_{\nu} = 0.61 \nonumber
\end{eqnarray}

\begin{figure}
\epsscale{0.7}
\plotone{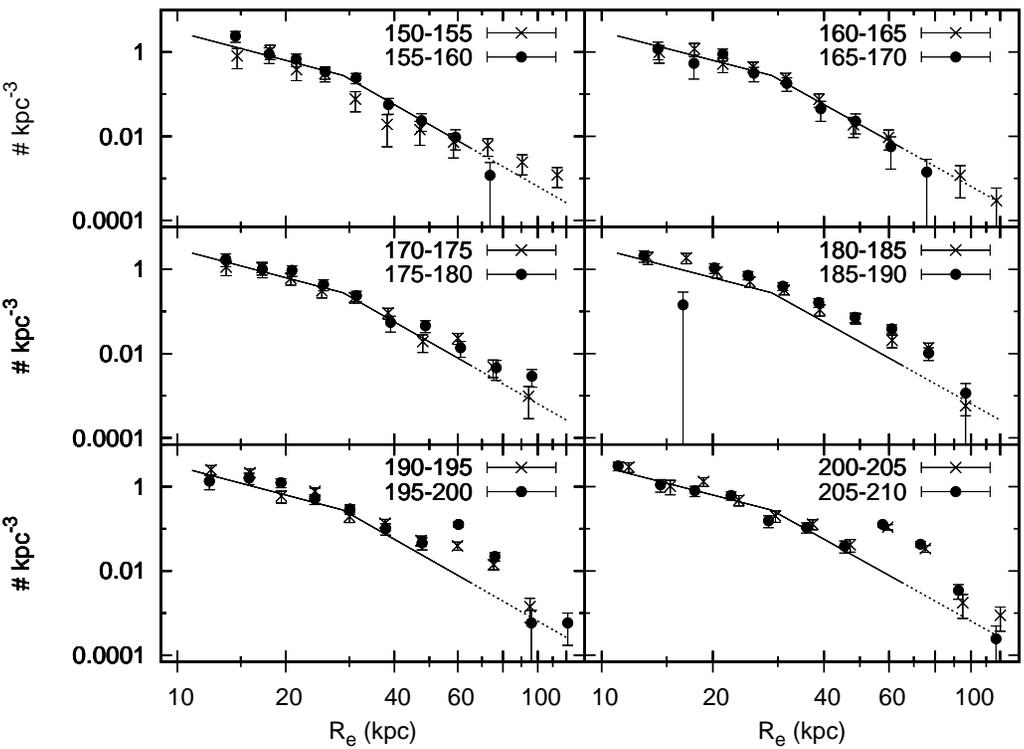}
\caption{The number density of RRLS is plotted against $R_e$ for
  different ranges in $\alpha$ (see text).  The lines are fits to the
  data in $150\degr \leq \alpha \leq 180\degr$.  They fit well the
  data except for the overdensities produced by the Sgr leading stream
  ($55\lesssim R_e \lesssim 90$ kpc, $180\degr \lesssim \alpha \lesssim
  210\degr$) and the Virgo Stellar Stream ($20\lesssim R_e \lesssim 35$
  kpc, $180\degr \lesssim \alpha \lesssim 190\degr$).}
\label{fig-denprof}
\end{figure}

The top panels in Figure~\ref{fig-denprof} shows that the fit of this
profile to the data in the range $150\degr \le \alpha \le 170\degr$,
is as good as the one obtained above with $R_{gc}$.  There is again
clear evidence for a break in the density profile.

The left middle panel shows that this profile is again somewhat too
steep beyond the break for the data in the range $170\degr \le \alpha
\le 180\degr$.  The right middle panel shows that in the range
$180\degr \le \alpha \le 190$, the profile underestimates the density,
but this is not surprising because this zone contains the VSS and the
Sgr stream.  The bottom panels show that in contrast to the spherical
profile, this profile fits well the data before the break.  The good
fit of this profile to the data before the break in all subdivisions
save the ones for $180\degr \le \alpha \le 190\degr$, where known halo
substructures exist, is evidence that the halo is flattened
and $q \sim 0.7$. Because of the presence of the Sgr Stream, our data
are not suitable for examining whether or not the flattening remains
the same beyond the break.

The profile that we find for the ellipsoidal halo is in reasonable
agreement with the results of \citet{sesar11}, who using a sample of
main-sequence stars found $q = 0.70\pm0.01$, a break at $27.8\pm0.8$ kpc,
and values of $-2.62\pm0.04$ and $-3.8\pm0.1$ for the slopes before
and after the break, respectively.  For the outer slope, our results
are in somewhat better agreement with results of \citet{deason11}, who
for a large sample of blue horizontal branch stars and blue stragglers
found $q \sim 0.6$, a break at $\sim 27$ kpc, and inner and outer
slopes of -2.3 and -4.6.  Recent investigations
\citep{watkins09,sesar11,akhter12} that have used RRLS as halo tracers
have also found breaks in the density profile with slopes similar to
what we find here.  The break radii found by \citet{watkins09} and
\citet{sesar10} for the RRLS in stripe 82 of the SDSS are similar to
what we have found here for a very different direction.
\citet{akhter12} found a break between 45 and 50 kpc in analysis of
the SEKBO RRLS survey.  Away from the Sgr stream (i.e. $\alpha <
180\degr$) where we can examine the halo slope at large distances, we
do not find evidence of a break at $45 - 50$ kpc (see
Figure~\ref{fig-denprof}). 

The result of using the ellipsoidal profile to estimate the background
RRLS as a function of $\alpha$ and $d_\sun$ is shown in
Figure~\ref{fig-Wedgecol}.  The density plus or minus from the profile
is given in terms of standard deviations, assuming Poisson statistics
and no uncertainty in the profile.  The large underdensity at $d_\sun
\lesssim 5 $ kpc is caused by the incompleteness due to CCD
saturation.  At $d_{\sun} \gtrsim 60$ kpc, the sample is becoming
incomplete (see above), which may explain the underdensity seen in
most directions at large $d_{\sun}$. The Sgr stream and the Sex dSph galaxy are
very prominent overdensities in Figure~\ref{fig-Wedgecol}.  There is also a
large but less extreme overdensity with rough boundaries of $170\degr
\lesssim \alpha \lesssim 205\degr$ and $8 \lesssim d_\sun \lesssim 20$
kpc.  At the largest $d_\sun$, this overdensity coincides with the VSS,
and at smaller distances, it may be part of the VOD.  It is discussed
in more detail below.

\subsection{Comparison with Sgr model by Law \& Majewski}

The Sgr stream is clearly seen in the LSQ data as a large overdensity 
at distances $>40$ kpc and $\alpha\gtrsim 180\degr$. The number of
RRLS belonging to the Sgr stream is large enough ($\sim 300$) that
detailed comparisons with models of the disruption of the Sgr galaxy
are possible.

In Figure~\ref{fig-Sgr} we show the footprint of the LSQ survey
together with the Sgr debris expected in the region according to the
recent model by \citet{law10a}. The color code of the Sgr debris is
the same as in \citet{law10a}, with the magenta points being particles
that became unbound in recent passages of Sgr and cyan and green
points corresponding to older debris. We will refer to them as young
and old debris respectively, although this has nothing to do with the
age of the stellar population they contain but with how long ago they
were separated from the main body of Sgr. This figure shows that while
there is considerable overlap in the directions to the young and old
debris, there are differences, particularly in their
distributions with $\alpha$.

To show the differences more clearly, separate plots are made in
Figure~\ref{fig-WedgeModel} for the model particles in the young and
old streams. Notice that the young and old streams roughly coincide
for $\alpha\gtrsim 180$ and $d_{\sun} > 30$ kpc, but the old stream
has a longer span in $\alpha$ than the young stream.  A simple
comparison with Figure~\ref{fig-Wedgecol} reveals a disparity between
the model particles for the old stream and the data.

In order to make a better comparison, we need to estimate the
normalization factor between the number of particles in the model and
the RRLS. To estimate this value, we chose the region encompassing
$200\degr<\alpha<210\degr$ and $0\degr<\delta<10\degr$, which has the
highest density of Sgr debris. In the range of distances from 40 to 60
kpc there are 546 particles in the model and 131 RRLS. Some of the
RRLS may, of course, be halo field stars. We integrated the density
profile determined above between 40 and 60 kpc, which yielded 8 for
expected number of halo RRLS. The remaining excess of 123 RRLS are
likely Sgr debris, which gives 4.4 for the ratio between the model
particles and RRLS.

The large size of this ratio suggests that LSQ survey contains
very few stars that belong to the old debris that crosses the survey
region at $\sim 10$ kpc (see Figure~\ref{fig-WedgeModel}). There are 66
  model particles in that nearby stream which means that it should
  contain $\sim15$ RRLS that are distributed along
  $\sim60\degr$ in $\alpha$. That is too low a density to be
  recognized over the general halo population, which is high at 10
  kpc.

There is a much better chance of detecting the old debris at greater
distances, where the density of the model particles is higher.  In
Figure~\ref{fig-Sgr_histo} we show histograms of the distances of the
RRLS and the model particles in 4 bins of $\alpha$.  The dotted lines
are the sum of the young and old of stream particles divided by 4.4,
and the shaded histograms are the young stream alone.  The solid lines
are the observed numbers of RRLS, after removing the expected number
of halo stars.  The two upper panels ($\alpha>190\degr$) show that the
debris is dominated by the young particles.  The RRLS distributions
match reasonably well both the total number of particles in the young
stream and the width of the stream along the line of sight in these
two panels.  They also follow the predicted gradient in mean distance
of the stream in the sense that the debris at
$190\degr<\alpha<200\degr$ is closer than at
$200\degr<\alpha<210\degr$.  There is, however, a systematic
difference between the distances of the RRLS and the model.  In the
uppermost panel the mean distance of the RRLS is 49 kpc while the mean
distance of the particles in the model is 51 kpc. This small
difference is probably due to a difference between the distance scales
for the RRLS and for the M giant stars to which the model is tied
\citep[see][]{law10a}.

Towards decreasing $\alpha$ (lower panels in
Figure~\ref{fig-Sgr_histo}), the old stream becomes more prominent, as
the number of young debris particles declines.  This produces the
differences between the dotted and shadded histograms.  Contrary to
the model, however, the excess of RRLS diminishes to the point of
being almost nonexistent at $170\degr<\alpha<180\degr$.  The old
stream is predicted by the model to contain 30 RRLS between 40 and 50
kpc, which should have been easily detected.  But there is essentially
no excess of RRLS in this region.  The predictions of the
\citet{law10a}'s model for the young stream are consistent with the
distribution of LSQ RRLS, but there is no sign of the older stream, at
least in this region of the sky.

\subsection{The overdensities in Virgo}

\begin{figure}
\epsscale{0.8}
\plotone{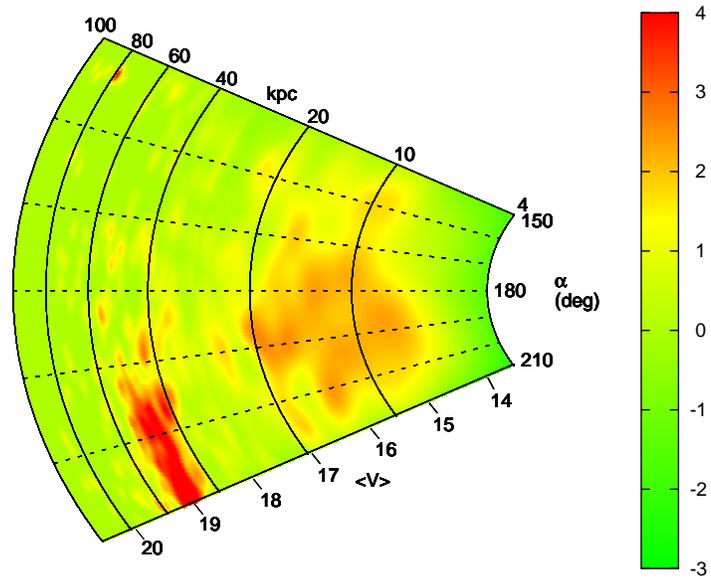}
\caption{The density distribution of the LSQ RRLS as a function of
  $\alpha$ and $d_{\sun}$.  The density scale is in units of standard
  deviations from the number density profile for the smooth halo
  (i.e. $\alpha \le 170\degr$).  The highest densities in the Sgr
  Stream are greater than 4 std. dev. above the background (see
  Figure~\ref{fig-denprof}), and therefore saturate on the density
  scale of this figure.  See the electronic edition of the Journal for
  a color version of this figure.}
\label{fig-Wedgecol}
\end{figure}

\begin{figure}
\epsscale{0.8}
\plotone{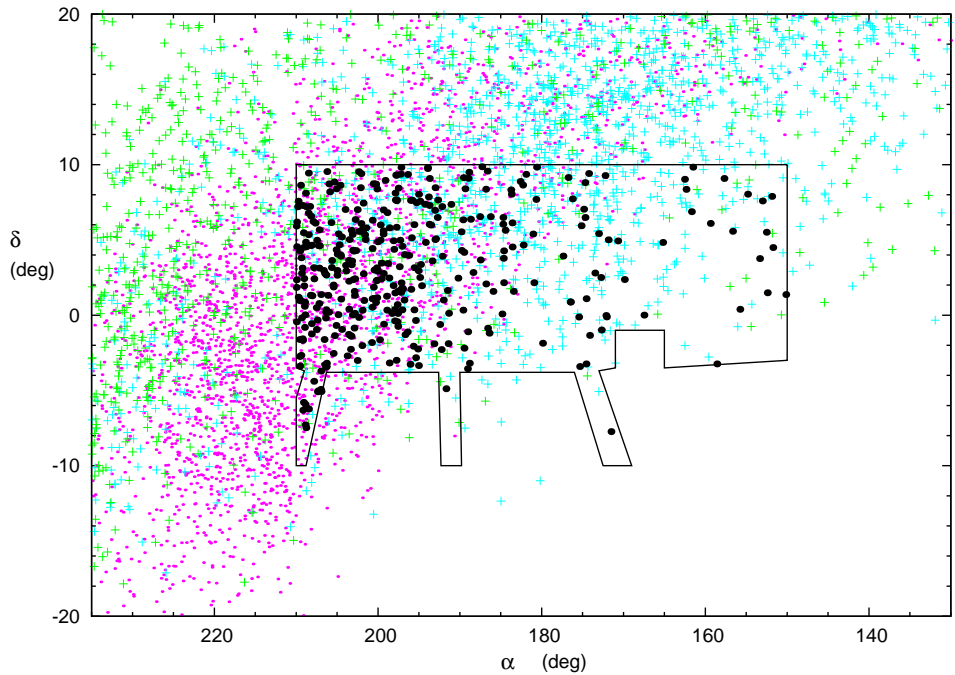}
\caption{Distribution on the sky of the particles (small dots and
  crosses) in the model of the disruption of the Sgr dSph galaxy by
  \citet{law10a}. To preserve clarity, only every other particle is
  plotted.  The dots are particles that were stripped during recent
  pericenter passages, while the crosses are particles that were
  stripped during older passages.  In the electronic edition of the
  Journal, the dots are colored magenta and the crosses are colored
  either cyan or green, following the same color scheme as
  \citet{law10a} for the epoch of stripping.  The footprint of the LSQ
  survey is shown by the polygon. RRLS in the range $40 \leq d_{\sun}
  \leq 60$ kpc are shown as black solid circles.  Note that they are
  concentrated where the younger stream particles (small dots) cross
  the footprint.  See the electronic edition of the Journal for a
  color version of this figure.}
\label{fig-Sgr}
\end{figure}

\begin{figure}
\epsscale{0.6}
\plotone{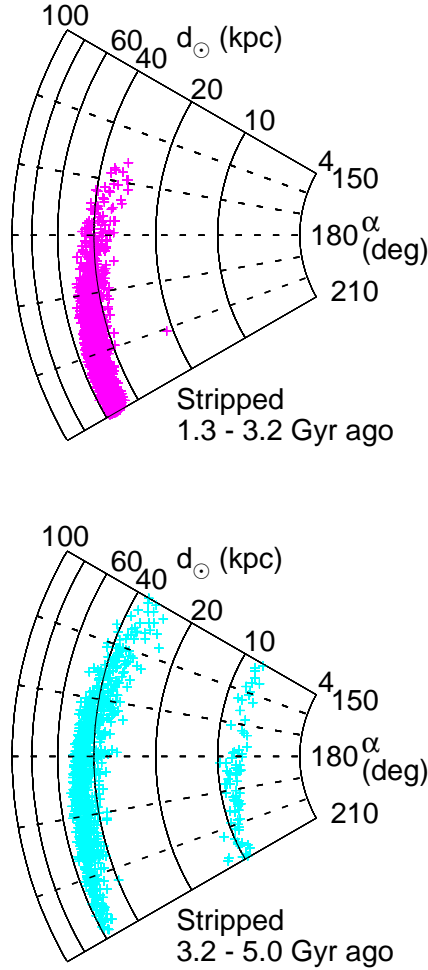}
\caption{The distribution in $\alpha$ and $d_{\sun}$ of the particles
  in the model that \citet{law10a} calculated for the stellar streams
  from the Sgr dSph galaxy.  The top and bottom diagrams show,
  respectively, the model stars that were separated from the main body
  of the galaxy 1.3---3.2 and 3.2---5.0 Gyrs ago.  See the electronic
  edition of the Journal for a color version of this figure, where the
  color scheme is the same as in Figure ~\ref{fig-Sgr} and in
  \citet{law10a}.}
\label{fig-WedgeModel}
\end{figure}

\begin{figure}
\epsscale{0.7}
\plotone{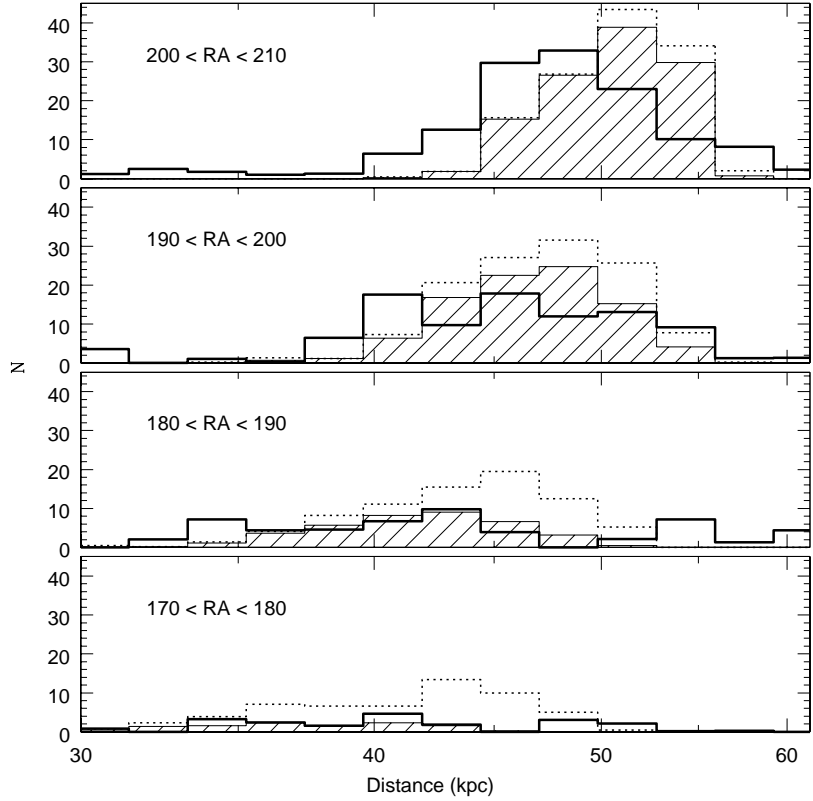}
\caption{Histograms of the distribution of distances of observed RRLS
  and of particles in the \citet{law10a} model of Sgr debris
  (normalized, see text) in 4 ranges of right ascension.  In each
  panel, the solid line is the number of observed RRLS, the shaded
  histogram is the number of young stream particles, and the dotted
  histrogram is the sum of both young and old stream particles.  In
  all 4 panels, the young debris particles appear to match the
  observed RRLS, although with an offset of about one bin width to
  larger distances, which is most evident in the upper panels.  Note
  the poor agreement in the lower two panels between the RRLS (solid
  line) and the sum of the young and old debris particles (dotted
  line), which suggests that the contribution from the old stream has
  been overestimated.}
\label{fig-Sgr_histo}
\end{figure}

As noted above, Figure~\ref{fig-Wedgecol} reveals a modest overdensity
with rough boundaries of $170\degr \lesssim \alpha \lesssim 205\degr$
and $8 \lesssim d_\sun \lesssim 20$ kpc.  In this general direction
and distance there have been several previous detections of
substructure using the spatial distributions and/or the radial
velocities of halo tracers
\citep{vivas01,newberg02,duffau06,newberg07,vivas08,juric08,keller10,walsh09,jerjen13}.
We will focus on the two overdensities that appear to be the best
documented, the Virgo Overdensity (VOD) and the Virgo Stellar Stream
(VSS), but this may be an over simplification since radial velocity
measurements suggest that this region may contain several more
substructures \citep{newberg07,vivas08,brink10,casey12,duffau13}.
Also note that some authors have argued that the VOD and the VSS are
different parts of the same overdensity, a conclusion that is hard to
rule out at this point.

The largest of the Virgo substructures is the Virgo Overdensity (VOD),
which is most clearly recognized as an excess of main-sequence stars
in the SDSS survey \citep{juric08,bonaca12b}.  The recent estimates by
\citet{bonaca12b} indicate it covers $\sim2000$ $\textrm{deg}^2$ of the sky and
possibly more, has a maximum density at $d_\sun \sim 11$ kpc, and is
detected at $d_\sun = 7$ kpc.  Compared to an axisymmetric region in
the southern galactic gap, \citet{bonaca12b} found an excess of 50\%
in the number density of main-sequence stars.
Figure~\ref{fig-Wedgecol} indicates that there is modest excess
of RRLS at $d_\sun \sim 11$ kpc from about $\alpha = 155\degr$ (l,b
$\sim 240\degr,50\degr$) to $205\degr$ (l,b $\sim 330\degr, 65\degr$).
According to Figures 2 and 3 in \citet{bonaca12b}, this region is
inside the VOD.  The region that we have used to estimate the number
density of RRLS in the smooth halo (i.e., $\alpha \le 170\degr$) also
overlaps with the VOD.  Consequently, the small excess of RRLS found
here may be a consequence of subtracting part of the VOD from itself.

\begin{figure}
\epsscale{1.0}
\plotone{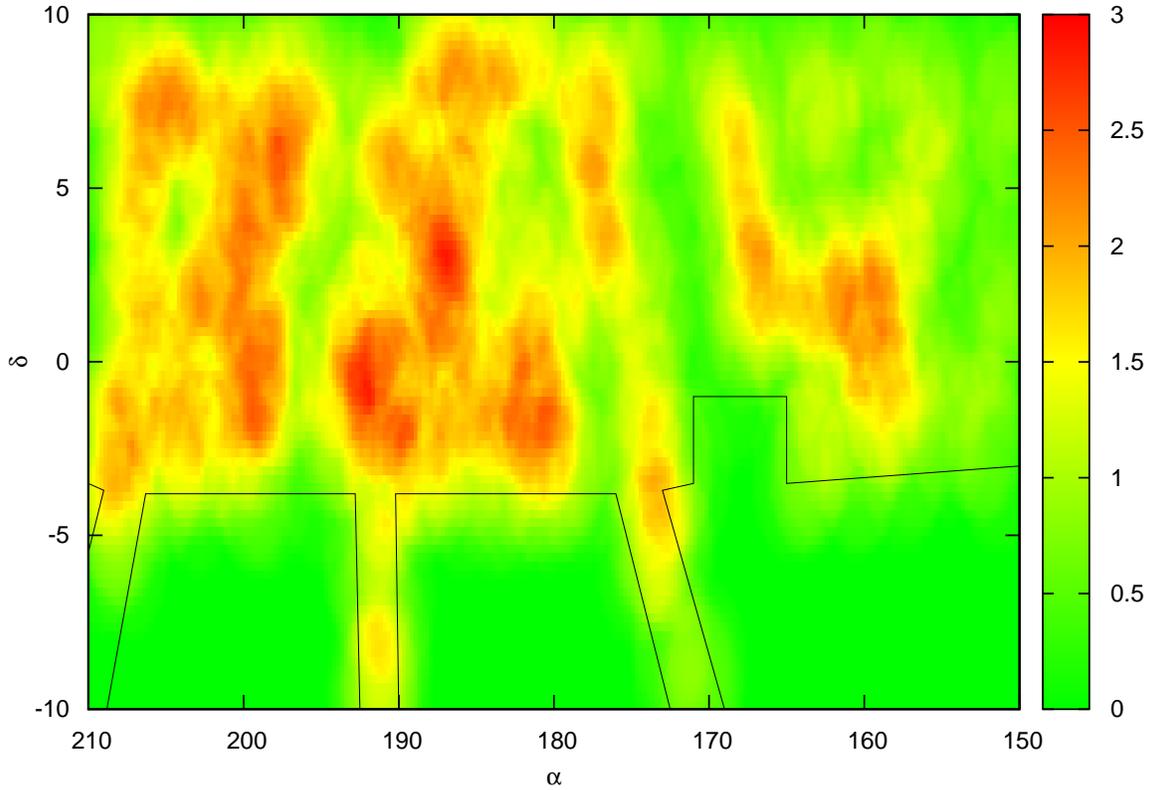}
\caption{The distribution on the sky of the density of RRLS in the
  range $17\le d_\sun \le 22$ kpc.  The density scale is the number of
  standard deviations from the density profile of the smooth halo.
  The solid line marks the boundary of the LSQ survey.  See the
  electronic edition of the Journal for a color version of this
  figure.}
\label{fig-VSSreg}
\end{figure}

Figure~\ref{fig-Wedgecol} shows that there is a more substantial
excess of RRLS at $d_\sun \sim 19$ kpc in the range $185\degr \le
\alpha \le 193\degr $, which is the location of the VSS
\citep[e.g.,][]{duffau06}.  The distribution in $\alpha$ and $\delta$
of the RRLS in the range $17 \le d_\sun \le 22 $ kpc is shown in
Figure~\ref{fig-VSSreg}.  In this plot, the VSS appears to divide into
three concentrations.  The two south of $\delta \sim 0$ are similar to
the distribution seen in the QUEST RRLS survey \citep{vivas06a}, and
they overlap with the detections of substructure by \citet{newberg02},
\citet{duffau06}, \citet{keller08}, and \citet{keller10} using
main-sequence turnoff stars, red giants, and subgiants.  They also
coincide with regions where radial velocity measurements
\citep{duffau06,newberg07,prior09a,starkenburg09,brink10,casey12,duffau13}
have revealed substructure in velocity space.  The feature at $\alpha
\sim 187\degr$ and $\delta \sim 3\degr$ appears to be the densest of
the three concentrations, but by only a small factor that may not be
significant.  The VSS appears to continue further to the north and may
reach $\delta \sim 10\degr$.  Figure~\ref{fig-VSSreg} shows a small
excess of RRLS at $\alpha \sim 191\degr$ and $\delta \sim -8\degr$,
which coincides with the area where \citet{newberg07} found evidence
of the VSS in velocity space.

In the same direction as the VOD and the VSS, \citet{walsh09} found
two small overdensities of stars using photometric data from the SDSS.
The most significant of these they dubbed Virgo Z, which they
considered to be a candidate UFD galaxy at $d_{\sun} \sim 40$ kpc.
Recently, \citet{jerjen13} constructed deep color-magnitude diagrams
(CMDs) at these positions, 1220-1 ($\alpha,\delta$) = ($185.077\degr,
-01.35\degr$)(Virgo Z in Walsh et al. 2009) and 1247-00
($191.992\degr,-00.75\degr$) and determined that they contain halo
substructures at similar distances, $d_{\sun} \sim 23$ kpc, and that
their stellar populations are similar in age, $\sim 8.2$ Gyr, and in
[Fe/H],$\sim -0.7$.  \citet{jerjen13} concluded that these
substructures are part of the same stellar stream and that it is separate
from the Sgr stream, which it resembles in age and [Fe/H], but not in
$d_{\sun}$.  The dominant stellar population of this stream is too
young to produce RRLS, but the same is true of the Sgr stream, which
is rich in RRLS that are part of its older, more metal-poor
population.  It is not clear that the stream identified by
\citet{jerjen13} is related to the large concentration of RRLS in this
direction, the VSS, because they differ in mean $d_{\sun}$ by $\sim 4$ kpc. 

\begin{figure}
\epsscale{1.0}
\plotone{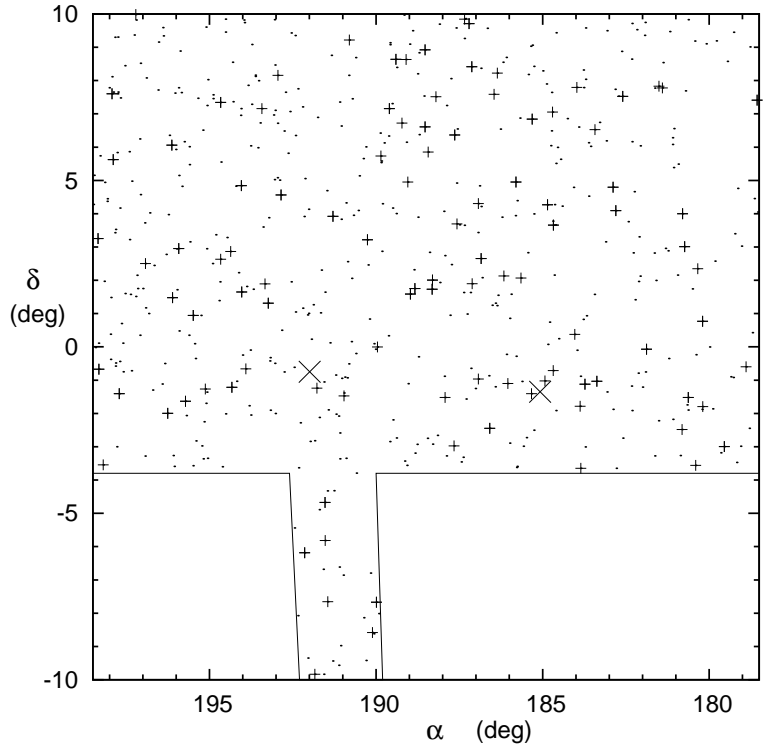}
\caption{The RRLS near the halo substructure identified by
  \citet{jerjen13}.  The large x's mark the positions of 1220-1
  (185.077, -01.35) and 1247-00 (191.992, -00.75).  The crosses are
  LSQ RRLS with $19.5 \le d_{\sun} \le 27.5$ kpc, which may coincide
  in distance with the substructure.  The small dots are LSQ RRLS that
  lie outside this distance interval.}
\label{fig-VirgoZ}
\end{figure}

To see if some LSQ RRLS may be associated with the substructure in
1220-1 and 1247-00, we have plotted in Figure~\ref{fig-VirgoZ} these
positions and the LSQ RRLS that are in the interval $19.5 \leq d_{\sun}
\leq 27.5$ kpc.  To arrive at this interval, we used the $1 \sigma $
extrema on the distances that \citet{jerjen13} measured at the two
positions and added in quadrature the 7\% uncertainty in the distances
of the RRLS that stem from the uncertainty in their absolute magnitudes
and from the photometric errors (see Vivas \& Zinn 2006).  While this
interval may seem large at first, it is important to recall that
streams have depth along the line of sight ($\sim 7$ kpc for the Sgr
stream, see Figure~\ref{fig-Sgr_histo}) and that the two positions
where this stream has been detected are separated by $\sim 7\degr$ and
may in fact have different $d_{\sun}$.

Figure~\ref{fig-VirgoZ} shows that while there is only one RRLS in the
above $d_{\sun}$ interval near field 1247-00 (LSQ726 at $d_{\sun} =
27.3$ kpc), there are are several near 1220-1.  The 4 RRLS within one
degree of the center of 1220-1 are listed in Table 4, where one can
see that 3 of the 4 have $d_{\sun} \sim 26.4$ kpc, which is within $1
\sigma $ of the $d_{\sun}$ of 1220-1 ($24.3\pm2.5$ kpc, Jerjen et al. 2013).
The two stars closest to the center of 1220-1 lie within the field
that is plotted in figure 10 of \citet{walsh09}, and they coincide
with regions of high stellar density in that figure.  While the 3
stars with similar $d_{\sun}$ in Table 4 appear to be the most likely
candidates for membership in this stream, it possible that the it
covers much of the area of Figure~\ref{fig-VirgoZ} and many more RRLS
may belong.

Figure~\ref{fig-VSSreg} shows that there are mild over densities at
$\alpha \sim 160\degr$ and $\delta \sim +2\degr$ and at $\alpha \sim
200\degr$, which may or may not be related to the VSS.  The one near
$\alpha = 200\degr$ coincides with an overdensity in subgiants that
was identified by \citet{keller10}.  Radial velocity measurements are
needed to test if these features are real and also the possibility
that they are related to the VOD, the VSS, or the stream detected by
\citet{jerjen13}.

\section{The Properties of the RR Lyrae Variables}

\subsection{The Oosterhoff Effect}

Much of the huge literature on the properties of RRLS has rightly
focussed on the Oosterhoff Effect \citep{oosterhoff39}, which can be
briefly summarized as follows.  The metal-poor ([Fe/H] $< -1.0$) MW
globular clusters (GCs) that contain five or more type ab variables
divide, with only a few exceptions, into two groups.  One group of
clusters (Oosterhoff Group I, OoI) have mean periods of the ab
variables $\langle P_{ab} \rangle$ near 0\fd55 and the other group
(OoII) have $\langle P_{ab} \rangle \sim 0\fd64$.  There are also
significant differences between the two groups in $\langle P_c
\rangle$ and in the ratio of the numbers of type c to the sum of all
types ($n_c/(n_{ab} + n_c)$) \citep[see Table 5, data
  from][]{smith95}.  The Oosterhoff dichotomy is produced by the
combination of metallicity and horizontal branch morphology
\citep[e.g.,][]{castellani03,catelan09}, and therefore, it is
complicated function of the metallicities, ages, and He abundances of
the clusters.

The Oosterhoff dichotomy is between the average properties of the RRLS
in the GCs.  In either the period-amplitude (P-A) diagram (sometimes
called the Bailey diagram) or the period histogram there is overlap
between the RRLS populating the OoI and OoII clusters.  In, for
example, the prototypical OoI cluster M3, $\sim 13 \%$ of the type ab
variables lie along the mean P-A relation of the prototypical OoII
cluster M15 \citep[see][]{cacciari05}, and a similar fraction of the
M15 variables lie along the P-A relation of M3
\citep[e.g.,][]{smith95,cacciari05,corwin08}.  In many OoI and OoII
clusters, there are RRLS that lie between the OoI and OoII P-A
relations \citep[e.g.,][]{cacciari05,
  szekely07,corwin08,zorotovic10,arellano11,dicriscienzo11}. Among the
MW GCs, only about 10\% have Oosterhoff intermediate (Oo-int)
properties, i.e.,$ 0\fd58\le \langle P_{ab} \rangle \le 0\fd62$, but the
frequency of Oo-int GCs is higher in the LMC and in the For and Sgr dSph
galaxies \citep{catelan09}.

It is not clear that the Oo-int GCs represent a physically distinct
class or are simply more even mixtures of OoI and OoII than typical
GCs \citep{smith09}.  Their P-A diagrams do not show clear evidence of
a sequence that is intermediate between the ones for the prototypical
OoI and OoII clusters M3 and M15 \citep[e.g.,][]{Kuehn11}.  The data
in \citet{catelan09} indicate that $\sim$30\% of the whole sample of
Oo-int clusters in the MW and its satellites are $\leq 0\fd005$ from the
boundaries in $\langle P_{ab} \rangle$ of the Oo-int region, which
suggests that the noise may have pushed some OoI and OoII clusters
into the Oo-int period range.  On the other hand, the majority of the
Oo-int clusters occupy a small area in a plot of [Fe/H] against HB
type \citep{catelan09}, which suggests that noise alone is not the
whole explanation.

The majority of the dSph galaxies of that are either satellites of the
MW or M31 \citep[e.g.,][]{pritzl05,kinemuchi08} and one UFD galaxy
\citep{garofalo13} are classified Oo-int on the basis of $\langle
P_{ab} \rangle$.  Their RRLS populations are typically mixtures of
stars that lie on the OoI and OoII sequences and in between
\citep[e.g.,][]{siegel00,held01,pritzl05,kinemuchi08}.  This is not
unexpected because these systems have metallicity spreads that overlap
with the [Fe/H] ranges of the OoI and OoII GCs.  Since many of these
systems have unusually red HB morphologies compared to GCs of the same
mean [Fe/H], HB morphology may be partially responsible as well.  The
best example of this may be the Draco dSph galaxy, whose red HB peters
out in the instability strip \citep{kinemuchi08}.

\subsection{The Oosterhoff Effect in subdivisions of the LSQ sample}

\begin{figure}
\epsscale{0.8}
\plotone{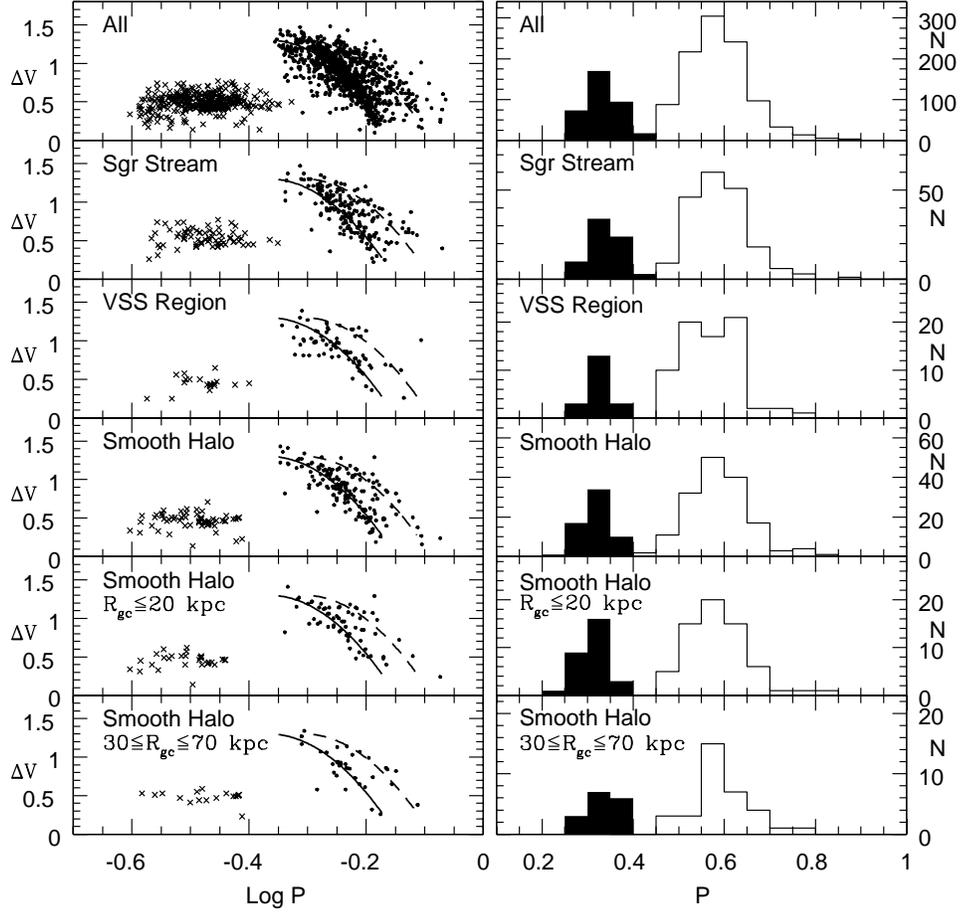}
\caption{The P-A diagrams (left) and the period histograms (right) for
  the whole sample of LSQ RRLS (minus the ones in the Sex dSph
  galaxy), for the Sgr stream, and for other subdivisions of the
  sample (see text).  The mean lines for the type ab variables in OoI
  and OoII globular clusters \citep{cacciari05} are plotted in the P-A
  diagrams as a solid and dashed curves, respectively.  The type c and
  type ab variables are plotted as x's and dots in the P-A diagrams
  and as filled and unfilled histograms, respectively.}
\label{fig-bailey}
\end{figure}

  The P-A diagram for the LSQ RRLS is shown in the upper left diagram
   of Figure~\ref{fig-bailey}, where one can see that it has the
   customary form \citep[see][]{smith95}.  Before plotting this
   diagram, we removed the stars that are probably members of the Sex
   dSph galaxy because they are not part of the halo field
   population. Figure ~\ref{fig-bailey} shows there are a small number of the
   type c variables that have very low amplitudes ($\Delta V < 0.2$), which
   raises some concern that they are not bona fide RRLS.  Since a few
   globular clusters \citep[e.g., NGC 2419][]{dicriscienzo11} and the
   LMC \citep{soszynski10} contain RRLS with similar properties, we
   have kept these stars in our catalog.

Following many previous workers, we compare the LSQ RRLS in the P-A
diagram with the mean loci for the type ab variables OoI and OoII
clusters \citep[from][]{cacciari05,zorotovic10}, but caution that the
amplitudes for some stars are probably affected by the Blazhko effect,
the cyclic variation in amplitude over periods of typically 20 to 200
days \citep[e.g.,][]{smith95}.  Studies of RRLS over long time base
lines have shown that $\ge 40\% $ the type ab and 5-40\% of the type c
variables \citep[e.g.,][]{kolenberg10} exhibit the Blazhko effect and
vary in amplitude by tenths of magnitude.  Since the LSQ observations
were obtained over only a few months, they are inadequate for
identifying stars that exhibit the Blazhko effect.  Consequently, the
amplitudes that we obtained may vary anywhere from the minimum to the
maximum of the Blazhko cycle, which will produce scatter in the P-A
diagram \citep[e.g.,][]{cacciari05}.  The period distributions LSQ
RRLS should be unaffected.  Since many of the investigations of the
RRLS in GCs and the dSph galaxies were also obtained over short
periods of time, some of the scatter in their P-A diagrams may also be
due to the Blazhko effect.

The type c RRLS in some OoI and OoII clusters follow different P-A
relationships
\citep[e.g.,][]{smith95,cacciari05,corwin08,zorotovic10}, but when
classifying by Oosterhoff type, most authors put more weight on the
type ab then the type c variables.  For example, \citet{zorotovic10}
classified the GC NGC 5286 as OoII, although nearly all of its type c
variables scatter around the P-A relation for OoI.  Because they
appear to be a less reliable indicator of Oosterhoff type, we do not
consider the P-A relationships of the type c variables in the
following.

Figure~\ref{fig-bailey} contains the P-A diagrams and the period
distributions for subdivisions of the LSQ sample
of particular interest (see also Table 5).  Every subdivision includes
the whole range in $\delta$ of the survey.  The one labelled Sgr
Stream is for the region defined by $ 193 \degr \le \alpha \le 210
\degr$ and $ 37 \le d_{\sun} \le 55$ kpc, which encompasses the densest
part of the leading stellar stream from the Sgr dSph galaxy.  The next
subdivision, $180 \degr \le \alpha \le 195 \degr $ and $15 \le d_{\sun}
\le 22$ kpc, is called the VSS Region because it contains the densest
part of the overdensity that is identified above with the VSS.  Note
that unlike the Sgr Stream, the contrast between the VSS
and the smooth halo background is not huge and contamination from
the background may be significant.  The next subdivision is the region,
$\alpha \le 170 \degr$, that we have used above to estimate the
density profile of the Smooth Halo.  The final two subdivisions
include the variables of the Smooth Halo inside and outside the break
in the density profile at $R_{gc} \sim 25$ kpc.  The region of the
break, where there may be overlap if two separate populations of RRLS
exist, has been purposely avoided.

For the subdivision of the LSQ variables, $\langle P_{ab} \rangle $
and $\langle P_{c} \rangle $, and the standard deviations of the
samples, $\sigma P_{ab}$ and $\sigma P_{c}$, and $n_c/(n_{ab}+n_c)$
are listed in Table 5.  The quantity OoII \% is defined as follows.
In the P-A diagram, we placed a curve of the same shape as the OoI and
OoII curves from \citet{zorotovic10} midway between the ones for OoI
and OoII and then determined the percentage of the total sample of
type ab RRLS that lie on the OoII side of this curve.  This procedure
gives an idea of the relative weighting of the OoI and OoII RRLS in
the sample.  It ignores the possibility that a distinct class of
Oo-int RRLS exists, which seems consistent with P-A diagrams of Oo-int
GCs.  Since the P-A diagram is affected by the Blazkho effect, we
caution that OoII \% may not a useful parameter to compare separate
studies that have different observational time spans.  This should not
be a factor here, where we use it to compare different samples of LSQ
RRLS.  The errors that are listed in Table 5 for $n_c/(n_{ab} + n_c)$
and OoII \% are based on Poisson statistics and are therefore lower
limits.

In each of the P-A diagrams of Figure~\ref{fig-bailey}, the type ab
variables scatter around the fiducials for OoI and OoII, with some
variables in between.  This and the values of $\langle P_{ab} \rangle
$, $\langle P_c \rangle$, $n_c/(n_{ab} + n_c)$, and OoII \% in Table 5
suggest that in each of these regions there is a mixture of OoI and
OoII that is weighted more towards OoI than OoII. 
 
On the basis of $\langle P_{ab} \rangle $, the VSS region is
classified as OoI, but it is close to the borderline.  The differences
between it and the Smooth Halo sample are minor, although the latter
is formally Oo-int.  The Kolmogorov-Smirnov (K-S) two sample test
indicates that there is no reason to reject the hypothesis that the
period distributions of the type ab variables in the VSS region and
the Smooth Halo have been drawn from the same parent distribution.
The value of $\langle P_{ab} \rangle $ found here for the VSS agrees
with the value, $\langle P_{ab} \rangle $ = 0\fd57, that is obtained
from 10 RRLS that are radial velocity members (Duffau et al. 2013).
Radial velocity measurements are needed for more stars before the
Oosterhoff properties of the VSS can be precisely determined.

In the Smooth Halo sample, there are no significant differences between
the samples of RRLS on either side of the break in the density
profile.  This suggests that there is not a large change across the
break that affect the properties of RRLS (e.g., metallicity and/or HB
morphology).

The mixture of OoI, OoII, and intermediate objects that we find here
for the Smooth Halo is similar to what is seen in the P-A diagrams of
other regions of the halo \citep{kinemuchi06,miceli08,szczygie09}, but
not all.  In the area studied by \citet{kinman12}, the fraction of OoII
variables is smaller than that found here, which suggests that the
properties of the Smooth Halo varies with position in the
Galaxy \citep[see also][]{miceli08}.

\subsection{The MW halo and its massive satellite galaxies}

According to the accretion scenario for the formation of the MW halo,
the RRLS in the outer halo should resemble the ones found in dwarf
galaxies.  Since the majority of the dSph satellite galaxies of both
the MW and M31 are Oo-int, several authors
\citep[e.g.,][]{catelan09,clementini10} have concluded that the MW
halo was unlikely to have formed from protogalactic fragments that
resembled the present-day dSph galaxies .  While the Smooth Halo
sample discussed above is formally Oo-int (see Table 4), its very near
the borderline with OoI, and it contains a larger fraction of OoI
variables than most dSph galaxies.  Because theoretical simulations of
satellite accretion suggest that much of the accreted halo comes from
the tidal destruction of a few massive satellites
\citep[e.g.,][]{bullock05,zolotov09}, it is of interest to compare the
RRLS populations of the MW halo and massive dwarf galaxies to see if
they provide a better match than the dSph galaxies in general.  Here
we compare the halo to the LMC and the SMC, and the two most luminous
dSph galaxies, Sgr and For, which are the only MW satellites that
contain GCs.  The mere presence of GCs in the outer halo suggests that
the accretion, if it is the origin of the halo, included at least a
few galaxies that were sufficiently massive to form their own GCs.
The accretion of a massive satellite, similar to the LMC, may be the
origin of the metal-rich tail on the metallicity distribution of the
halo \citep{frebel10}.  Other dwarf galaxies in the Local Group are
known to contain both GCs and RRLS, and they too may be analogues of
the massive satellites that have been accreted.  Unfortunately, the
existing surveys of their RRLS populations are in small fields,
typically at the outskirts of the galaxies, and therefore, they may
not be representative of the whole galaxies \citep[e.g., M32,
  see][]{sarajedini12}.  For the comparisons that we make here, it is
important to note that the estimates of the total population of RRLS in
For range from $\sim 660$ \citep{vivas06a}, which is based on the
sample of \citet{bersier02} and consequently may be too low
\citep[see][]{greco05}, to $\sim 2000$ \citep{clementini10}, and that there
are several thousands of RRLS already measured in each of the other
galaxies.  The accretion of just one galaxy that resembled any one of
these four would put as many, and possibly several times more RRLS in
the halo than all the other known dSph and UFD galaxies \emph{combined}
\citep[see][]{clementini10}.  Each of the 4 galaxies has experienced
long histories of star and cluster formation, which in the cases of
the LMC and the SMC continue to the present time.  Since we are
considering here only the RRLS, which are very old, these galaxies may
be used as analogues of galaxies that merged with the MW a few billion
years after the Big Bang, a time when much of the accreted halo formed
according to the theoretical simulations \citep{bullock05,zolotov09}.
We have ignored the putative dwarf galaxy in Canis Major in this
discussion because it is not certain that it even exists and because
if it does, it contains very few RRLS \citep{mateu09}.

In Table 5, we have collected data for the LMC and the SMC from the
OGLE project \citep{soszynski09,soszynski10}.  Because we are unable
to separate type d and type e RRLS from type c in the LSQ survey, the
results in Table 5 were obtained by lumping together these types for
the LMC and SMC.  The sample of Sgr RRLS is from
\citet{cseresnjes01}, and to limit the possible contamination from RRLS in
the galactic bulge or halo, we limited this sample to RRLS in the more
centrally located of his two fields and to ones fainter than $B_{i}$ =
18.6 \citep[see][]{cseresnjes00}.  The type d variables in this sample
are counted as type c here.  The For data are from \citet{bersier02}
who did not use the P-A diagram to decide pulsation type, but instead
divided the stars into type c and type ab depending on whether their
periods were to the short or long period side of 0.47 day.  Bersier
and Wood noted that they rejected many possible RRLS in order to
obtain a clean sample of RRLS instead of a more complete one.  Since
\citet{greco05} found similar results for $\langle P_{ab} \rangle $
from another, although smaller, sample of For RRLS, the mean
properties of the Bersier and Wood sample are probably representative
of the galaxy.

\begin{figure}
\epsscale{0.7}
\plotone{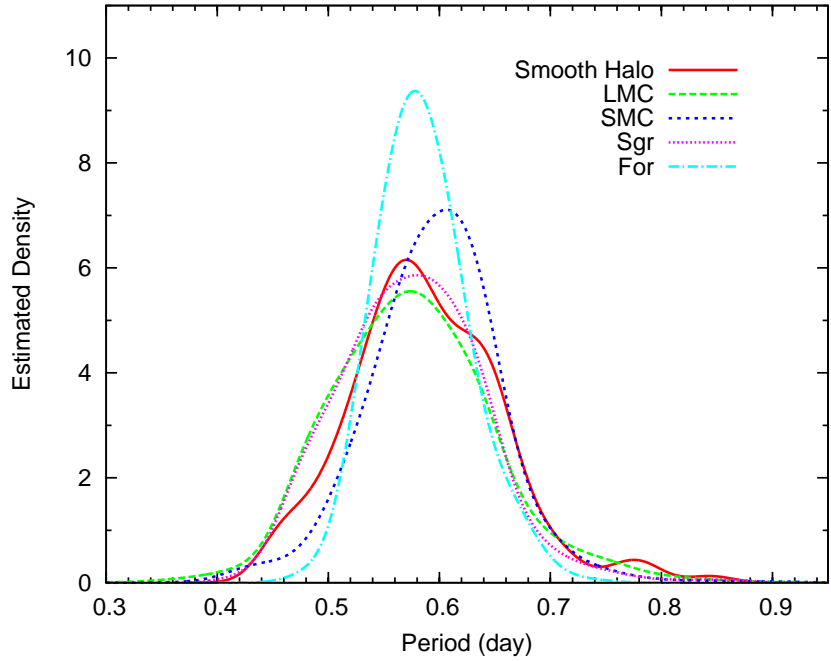}
\caption{The estimated probability density functions for the periods
  of the type ab variables in the Smooth Halo, LMC, SMC, Sagittarius
  (main-body), and Fornax are compared.  These functions were
  calculated using a Gaussian kernal with a standard deviation of
  $0\fd02$.  See the electronic edition of the Journal for a color
  version of this figure.}
\label{fig-satden} 
\end{figure}

On the basis of the values of $\langle P_{ab} \rangle$ in Table 5, the
LMC and Sgr are OoI and the SMC and For are Oo-int.  Note, however, if
we had adopted the results that the MACHO project had obtained for the
LMC \citep{alcock96}, it would be classed as Oo-int, albeit near the
OoI borderline, which underscores that different samples, although
large in numbers of variables, can lead to different classifications
for the same object.  In terms of $\langle P_{ab} \rangle$, every
galaxy in Table 5 lies near the boundary between OoI and Oo-int.  The
values of $\langle P_{c} \rangle $ and $n_c/(n_{ab} + n_c)$ are
consistent with OoI or Oo-int, but not OoII. The period distributions
of the type ab variables in these galaxies are compared with the one
for the Smooth Halo in Figure~\ref{fig-satden}, which reveals some
interesting similarities and differences.  As noted by
\citet{cseresnjes01}, the period distributions of the LMC and Sgr are
strikingly similar.  Large differences exist between the distributions
of the LMC and the SMC \citep{soszynski10}, particularly at the
shorter periods.  Compared to the other galaxies, For has a
particularly narrow distribution, but one wonders if this may be due
in part to the incompleteness of the sample \citep[see][]{bersier02,
  greco05}.  We selected only the Smooth Halo sample to compare these
distributions, because the other regions of the LSQ survey contain
contributions from either the Sgr Stream or the VSS.
Figure~\ref{fig-satden} shows that the Smooth Halo sample has a broad
period distribution that resembles more the distributions of Sgr and
the LMC than the ones of the other two galaxies.  While there is no
reason to believe that the Smooth Halo originated from the accretion
of just one galaxy, on statistical grounds (i.e., the K-S test at 0.05
significance) there is no reason to reject the hypothesis that the
Smooth Halo has the same period distributions as the LMC or Sgr.

\begin{figure}
\epsscale{1.0}
\plotone{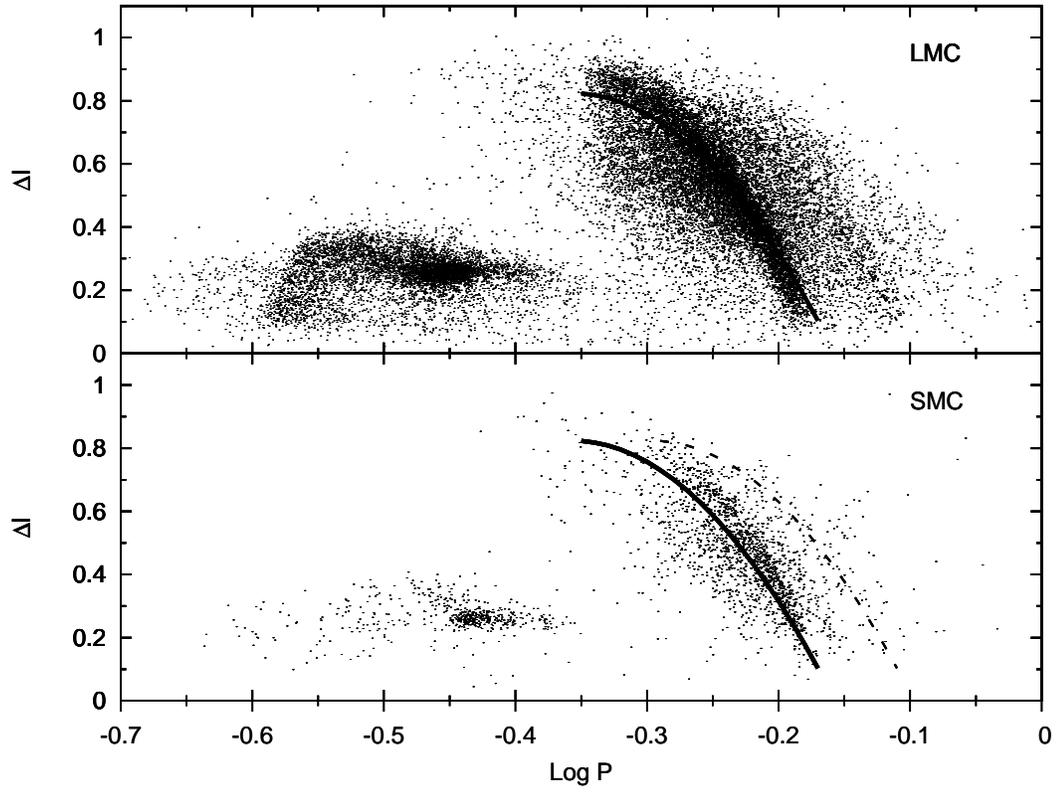}
\caption{The Period-Amplitude diagrams for the RRLS observed by the
  OGLE project in the LMC and the SMC.  The I-band amplitudes, $\Delta
  I$, of the OoI and OoII curves, solid and dashed lines respectively,
  were transformed from the V-band to the I-band (see text).}
\label{fig-LMCSMC}
\end{figure}

The P-A diagram for Sgr \citep[see fig. 8 in][]{cseresnjes01} shows
that a large fraction of its type ab variables scatter around the OoI
relation.  The P-A diagrams of the LMC and the SMC samples are shown
in Figure~\ref{fig-LMCSMC}.  Because the OGLE project measured the
amplitudes in the Cousin's I-band \citep{soszynski09}, it was
necessary to transform the V-band P-A relationships for OoI and OoII
\citep{zorotovic10} to the I-band.  To do this, we used equation 1 in
\citet{dorfi99}, which is based on the V- and I-band amplitudes of
RRLS in several GCs.  Figure~\ref{fig-LMCSMC} shows that in both the
LMC and the SMC, there are large concentrations of type ab variables
to left and slightly to the right of the OoI curve, and relatively few
variables in the vicinity of the OoII curve.  In the LMC diagram,
there is a dense band of type ab stars, which at large amplitudes lies
to the right of the OoI curve, but at lower amplitudes coincides with
the curve or to the left of it.  It is not clear if this is a sign
that the curve, which is based on many fewer RRLS in the prototypical
OoI GC M3 \citep{cacciari05}, does not accurately trace the OoI
sequence, or if the LMC has an strong Oo-int component that manifests
itself at only the largest amplitudes (i.e., $\Delta I \geq 0.6$
mag). In either case, it is clear that the accretion of galaxies similar to
the LMC or the SMC would have heavily weighted the halo toward OoI.
Because \citet{bersier02} did not measure the amplitudes of the RRLS,
it is not possible to plot a P-A diagram for the For sample.  It is
unlikely to contain a large population of OoII variables because only
2\% of its type ab variables have P $\geq 0\fd68$ day, whereas OoII GCs
typically have larger percentages (e.g., 26\% in the case of the OoII
GC NGC 2419, Di Criscienzo et al. 2011).  The P-A diagram that
\citet{greco05} plotted for their sample of 110 RRLS in For clearly shows
that the majority of the type ab variables follow the OoI relation.

The accretion of galaxies similar to the ones in Table 5 may explain
the preponderance of OoI variables in the MW halo.  The accretion of
galaxies similar to the other dSph galaxies, mostly Oo-int, and the
UFDs, mostly OoII, is also expected if the accretion hypothesis is
correct.  In the period distribution and the P-A diagram, the RRLS
from these other sources will largely blend in with ones from any
massive satellites, and the relative contributions of these different
types of systems may be difficult to disentangle.  Because the
existing high mass satellites contain small fractions of stars near
the OoII sequence, the low-mass satellites may be more important
source for OoII than for OoI stars.

If the galaxies in Table 5 are truly analogues of the galaxies that
were accreted, then their GC populations should also resemble that of
the outer halo.  \citet{mackey04,vandenBergh04,mackey05} have compared
the GCs in these galaxies with the ones in the outer halo of the MW,
and found substantial overlap in terms of position in the HB type -
[Fe/H] plane, luminosity, and structural parameters such as half-light
radius and core radius.  They also noted that in terms of these
parameters the outer halo GCs are distinctly different from the inner
halo and bulge/disk GCs.  They concluded that the GCs of the outer MW
halo are likely to have been accreted from dwarf galaxies resembling
the LMC, SMC, and the Sgr and For dSph galaxies.  \citet{catelan09}
has argued, however, that the GCs in the LMC, For and Sgr are not a
good match to the ones in the MW because several of them lie in a
triangular area of the HB type - [Fe/H] plane that is unoccupied by MW
GCs.  This small region appears to be largely responsible for the Oo-int
phenomenon, and therefore it may be the origin of the larger
frequencies of Oo-int GCs in these dwarf galaxies than in the MW halo
\citep{catelan09}.  To occupy this region a GC must have [Fe/H] and
age within narrow ranges \citep[see fig. 7][]{catelan09};
consequently, it is unclear that all dwarf galaxies must necessarily
contain such clusters and have greater frequencies of Oo-int clusters
than the outer halo of the MW.  A good example of the vagaries of GC
formation in dwarf galaxies is provided by the SMC. Despite its large
population of very old stars, e.g., the RRLS, the SMC contains just
one old GC, NGC 121 \citep[e.g.,][]{piatti07}, which contains too few
RRLS to be classified by Oosterhoff type.  In terms of position in the
HB type - [Fe/H] plane, luminosity, and structural parameters, it is
similar to the GCs in the outer halo of the MW
\citep{mackey04,vandenBergh04,mackey05}, but it does not lie in the
triangular area associated with the Oo-int phenomenon.

 The hypothesis that the accretion of satellite galaxies was a major
 contributor to the MW halo does not require that an exact match exist
 between the halo and the still existing satellites, just that the
 properties of the halo be consistent with a physically reasonable
 mixture of the old stellar populations and the GC populations of
 extant satellites in general \citep[e.g.,][]{zinn93,mackey04}.  The
 papers by Mackey, van den Bergh and Gilmore (see above) suggest that
 this is the case for the outer halo GCs.  The comparisons that we
 have made here suggest that the RRLS population of the outer halo may
 be explained by satellite accretion, if it included some relatively
 massive systems.

\subsection{The RRLS populations of the Sgr Stream and its main body}

\begin{figure}
\epsscale{0.8}
\plotone{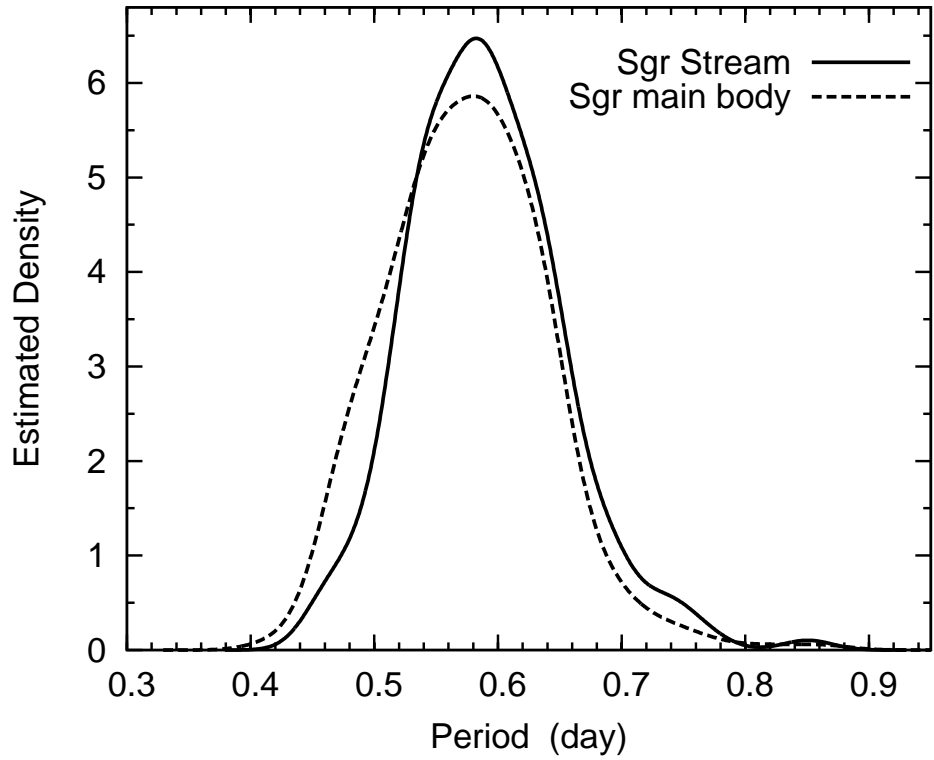}
\caption{The solid and dashed lines are, respectively, the estimated
  probability density functions for the periods of the type ab RRLS in
  the Sgr dSph galaxy and for the LSQ RRLS in the Sgr Stream, which
  were calculated using Gaussian kernal with a standard deviation of
  $0\fd02$.}
\label{fig-estden}
\end{figure}

One case where one might expect an exact match of RRLS populations is
between the Sgr Stream and its main body, and this appears not to be
the case.  The data in Table 5 show that a difference exists between
them in $\langle P_{ab} \rangle$.  We have examined this further by
using the K-S two sample test to see if their distributions in
$P_{ab}$ are indeed different. With the LSQ sample for the Sgr Stream
and the sample described above from \citet{cseresnjes01} for the main
body, the K-S test indicates that the null hypothesis that they were
drawn from the same parent distribution can be rejected at a
significance level $< 0.025$.  The estimated probability density
functions of the two samples are compared in Figure~\ref{fig-estden},
where one can see that the major difference is that the main-body
sample has a higher percentage ($\sim 20\% $) of variables
with $P_{ab} \leq 0.52$ than the stream ($\sim 10\%$).  Because
\citet{cseresnjes01} did not find a significant gradient in RRLS
properties across the main-body, this difference appears to be limited
to one between the outskirts of the pre-disturbed galaxy, which
presumably now make up the Sgr streams, and its central regions.

Among type ab RRLS, there is a rough, inverse correlation between
period and [Fe/H].  According to the relationship that
\citet{sarajedini06} determined from the comparison of the periods
and spectroscopically determined [Fe/H] values for 132 field type ab
variables, the period of 0.52 day corresponds to [Fe/H] $\sim -1.2$.
The difference between the period distributions in
Figure~\ref{fig-estden} may be a sign that the [Fe/H] distribution of
the main-body has a more extensive metal-rich tail, $\lesssim -1.2$,
than does the stream.  Since other studies have found evidence for
chemical abundance differences along the Sgr streams and/or between
the main-body and the streams \citep{chou07,monaco07,chou10,shi12},
this explanation seems most likely, although alternatives, such as a
difference in distribution of stars accross the instability strip
cannot be ruled out.

\section{Summary}

Our survey of $\sim 840$ $\textrm{deg}^2$ of the sky has yielded light
curves, periods, and mean magnitudes for 1013 type ab and 359 type c
RRLS.  The completeness of the survey is estimated to be $\gtrsim
70\%$ over the range $8\lesssim d_{\sun} \lesssim 60$ kpc and decline
to zero at $d_{\sun} \sim 100$ kpc.  The survey region overlaps with
part of the Sex dSph galaxy, and 6 new RRLS and 3 new AC
variables were discovered in this galaxy.

In directions away from the prominent halo substructures, the spatial
distribution of the LSQ RRLS is consistent with an ellipsoidal model
that is flattened towards the galactic plane.  A ratio of minor to
major axes of 0.7, as estimated by \citet{sesar11}, provides a good
match to our RRLS data.  The number density profile of the RRLS has a
break in its power-law slope at $R_{gc} \sim 25$ kpc, in the sense
that it is clearly steeper at larger distances.  Since other
investigators have seen similar breaks in different directions, at
approximately the same $R_{gc}$, and using a variety of halo tracers
(RRLS, main-sequence stars and blue horizontal branch stars), this
break appears to be a general feature of the MW halo.  We do not find
evidence for a change in $\langle P_{ab}\rangle$, $\langle P_c
\rangle$, $n_c/(n_{ab} + n_c)$, or OoII \% across the break.  It is
important to examine this again once larger samples of RRLS are
available.

The most prominent substructure in the survey region is the leading
stream from the Sgr dSph galaxy, which contains $\sim 20\%$ of the
whole sample of RRLS.  In the direction and distance from the Sun, the
stream outlined by the RRLS is consistent with the predictions of the
recent model by \citet{law10a} for stars that were stripped from the
galaxy between 1.3 and 3.2 Gyr ago.  According to this model, stars
that were stripped from the galaxy between 3.2 and 5.0 Gyrs have a
broader distribution in $\alpha$ than the more recently stripped
stars.  Our observations do not agree with this prediction for the
older stream stars.  Our failure to detect this older stream may mean
that it contains far fewer stars than predicted by the model, is
displaced out our survey region, or is coincident in direction and
distance with the younger stream. 

The Sgr stream has a lower percentage of short period type ab
variables than the main-body of the galaxy.  One possible
interpretation is that main-body of Sgr has a more extensive
metal-rich tail to its [Fe/H] distribution than the stream.  This is
consistent with other evidence that suggests that metallicity
differences exist between the main-body of Sgr and its streams.

The second most prominent over density is related to the previously
identified substructures in Virgo, the VOD and the VSS.  The densest
part of this feature coincides in direction and distance, $\sim 20$
kpc with the VSS.  Our survey provides evidence that the VSS extends
to the north of previous detections, i.e., $\delta \gtrsim 2\degr $.
In our data, the VSS is the most distant part of a substructure that
extends over the range $8 \lesssim d_\sun \lesssim 20$ kpc, which on
the near side overlaps with the much larger VOD
\citep[see]{juric08,bonaca12b}.  We caution that this apparent
connection in space between the VSS and VOD may not mean a common
origin.  Radial velocity measurements \citep[see][]{duffau13} reveal
that there are several different clusterings in velocity and distance
that are more indicative of several separate moving groups rather than
one large one.  It is not clear at this time, if any of these groups
are related to the substructure that \citet{jerjen13} have recently
discovered in the CMDs of the fields 1220-1 and 1247-00, which is in
the same direction but appears to be more distant at $d_{\sun} \sim
23$ kpc.  Some LSQ RRLS that lie close to these positions have values
of $d_{\sun}$, which suggest possible membership in this new stellar
stream.

The RRLS in the Sgr leading stream, the VSS, and the smooth halo are
similar to each other in $\langle P_{ab}\rangle$, $\langle P_c
\rangle$, $n_c/(n_{ab} + n_c)$, and OoII \%.  Their values of $\langle
P_{ab}\rangle$ place them near the dividing line between OoI and
Oo-int.  In the P-A diagram, their type ab variables resemble a
mixture of OoI and OoII that is more heavily weighted towards OoI
($\sim 27\%$ OoII).  These samples of RRLS in the MW halo resemble the
ones found in the field populations of the 4 most massive satellite
galaxies of the MW.  The similarities between
the MW halo and these satellite galaxies are consistent with
theoretical results, which suggest that the accretion of a few massive
satellite galaxies produces a large fraction of a galaxy's halo.

\acknowledgments

We thank the anonymous referee whose comments on the manuscript have led to
several improvements.

This research has been supported by NSF grant AST-1108948 and DOE
grant DE FG0 ER92 40704 to Yale University and by the Provost Office
of Yale University.  This project would not have been possible without
the public release of the data from the Sloan Digital Sky Survey III.
Funding for SDSS-III has been provided by the Alfred P. Sloan
Foundation, the Participating Institutions, the National Science
Foundation, and the U.S. Department of Energy Office of Science. The
SDSS-III web site is http://www.sdss3.org/.

SDSS-III is managed by the Astrophysical Research Consortium for the
Participating Institutions of the SDSS-III Collaboration including the
University of Arizona, the Brazilian Participation Group, Brookhaven
National Laboratory, University of Cambridge, Carnegie Mellon
University, University of Florida, the French Participation Group, the
German Participation Group, Harvard University, the Instituto de
Astrofisica de Canarias, the Michigan State/Notre Dame/JINA
Participation Group, Johns Hopkins University, Lawrence Berkeley
National Laboratory, Max Planck Institute for Astrophysics, Max Planck
Institute for Extraterrestrial Physics, New Mexico State University,
New York University, Ohio State University, Pennsylvania State
University, University of Portsmouth, Princeton University, the
Spanish Participation Group, University of Tokyo, University of Utah,
Vanderbilt University, University of Virginia, University of
Washington, and Yale University.



\clearpage

\clearpage

\begin{deluxetable}{ccccccccccc}

\rotate

\tabletypesize{\footnotesize}


\tablecaption{RR Lyrae Stars}

\tablenum{1}


\tablehead{\colhead{LSQ} & \colhead{$\alpha_{2000}$} & \colhead{$\delta_{2000}$} &
  \colhead{$N_{obs}$\tablenotemark{a}} & \colhead{$\langle V \rangle$} & \colhead{Amp} &
  \colhead{P} & \colhead{HJD\tablenotemark{b}} & \colhead{Type} &
  \colhead{Flag\tablenotemark{c}} & \colhead{Note\tablenotemark{d}}
  \\ \colhead{} & \colhead{(deg)} & \colhead{(deg)} & \colhead{} &
  \colhead{(mag)} & \colhead{(mag)} & \colhead{(day)} &
  \colhead{(day)} & \colhead{} & \colhead{} & \colhead{} }

\startdata
1 & 150.1707 & 1.3682 & 46 & 18.79 & 1.17 & 0.48960 & 55264.51565 & ab & 0 & \nodata \\
2 & 150.3709 & 0.1833 & 55 & 16.47 & 1.14 & 0.60535 & 55266.57127 & ab & 0 & \nodata \\
3 & 150.3886 & 1.7246 & 51 & 15.07 & 1.06 & 0.54243 & 55277.56310 & ab & 0 & \nodata\\
4 & 150.4103 & 0.2355 & 55 & 18.58 & 0.44 & 0.33643 & 55279.60525 &  c & 0 & \nodata \\
5 & 150.4330 & 8.6984 & 40 & 17.03 & 0.50 & 0.29651 & 55273.52587 &  c & 0 & \nodata \\
\enddata


\tablenotetext{a}{Number of observations used in template fit}
\tablenotetext{b}{Heliocentric Julian Date of Maximum Brightness}
\tablenotetext{c}{Calibration: 0 = good, 1 = poor}
\tablenotetext{d}{Suspected member of the Sex dSph galaxy = Sex dSph;
  if available V number from \citet{mateo95}} 
\tablecomments{Table 1 is published
in its entirety in the electronic edition of The Astrophysical
Journal.  A portion is shown here for guidance regarding its form and
content.}

\end{deluxetable}

\begin{deluxetable}{cccccccccc}

\rotate

\tabletypesize{\small}


\tablecaption{Anomalous Cepheids in the Sextans dSph galaxy}

\tablenum{2}

\tablehead{\colhead{LSQ} & \colhead{$\alpha_{2000}$} & \colhead{$\delta_{2000}$} & \colhead{$N_{obs}$} & \colhead{$\langle V \rangle$} & \colhead{Amp} & \colhead{P} & \colhead{HJD\tablenotemark{a}} & \colhead{$M_V$} & \colhead{Note} \\ 
\colhead{} & \colhead{(deg)} & \colhead{(deg)} & \colhead{} & \colhead{(mag)} & \colhead{(mag)} & \colhead{(day)} & \colhead{(day)} & \colhead{(mag)} & \colhead{} } 

\startdata
24 & 152.9940 & -2.0379 & 47 & 19.98 & 0.48 & 0.45917 & 55280.54211 & +0.22 & \nodata \\
31 & 153.2278 & -1.5302 & 53 & 19.37 & 0.43 & 0.49278 & 55274.54471 & -0.39 & V19 \\
34 & 153.3578 & -1.6825 & 44 & 19.89 & 0.30 & 0.50640 & 55273.63399 & +0.13 & V34 \\
36 & 153.4017 & -1.6314 & 51 & 18.88 & 0.89 & 0.69300 & 55286.51409 & -0.88 & V1 \\
40 & 153.6185 & -1.6865 & 50 & 18.64 & 0.60 & 2.02186 & 55281.52954 & -1.12 & Q119 \\
55 & 154.9881 & -2.0505 & 46 & 19.55 & 0.88 & 0.42806 & 55278.55108 & -0.21 & \nodata \\
\enddata

\tablenotetext{a}{Heliocentric Julian Date of maximum brightness - 2400000.0}
\tablecomments{V and Q numbers from Mateo et al. (1995) and Vivas et al. (2004), respectively.}


\end{deluxetable}

\begin{deluxetable}{cccc}




\tablecaption{The Observations}

\tablenum{3}

\tablehead{\colhead{LSQ} & \colhead{HJD\tablenotemark{a}} & \colhead{V} & \colhead{std. dev.\tablenotemark{b}} \\ 
\colhead{} & \colhead{(day)} & \colhead{(mag)} & \colhead{(mag)} } 

\startdata
     1 &   55286.515908 &   17.960 &   0.024\\ 
     1 &   55261.565315 &   17.920 &   0.011\\ 
     1 &   55260.600868 &   17.958 &   0.013\\ 
     1 &   55262.587966 &   18.167 &   0.012\\ 
     1 &   55286.558406 &   18.294 &   0.032\\ 
\enddata


\tablenotetext{a}{Heliocentric Julian Date - 2400000.0}  
\tablenotetext{b}{Based on photon statistics}  
\tablecomments{Table 3 is published in its entirety in the electronic edition of The Astrophysical Journal.  A portion is shown here for guidance regarding its form and content.}


\end{deluxetable}

\begin{deluxetable}{cccccccc}




\tablecaption{RR Lyrae Variables near 1220-1}

\tablenum{4}
 
\tablehead{\colhead{LSQ} & \colhead{$\alpha_{2000}$} &
  \colhead{$\delta_{2000}$} & \colhead{Type} & \colhead{$\langle V
    \rangle$} & \colhead{$\langle V \rangle_{0}$} & \colhead{$d_{\sun}$} &
  \colhead{$\phi$\tablenotemark{a}} \\ \colhead{} &
  \colhead{(deg)} & \colhead{(deg)} & \colhead{} & \colhead{(mag)} &
 \colhead{(mag)} &  \colhead{(kpc)} & \colhead{(deg)} }

\startdata
515 & 184.67940 & -0.71070 & c & 17.18 & 17.07 & 20.2 & 0.754 \\
525 & 184.92920 & -1.02490 & c & 17.73 & 17.63 & 26.0 & 0.358 \\
532 & 185.33610 & -1.40380 & ab & 17.77 & 17.66 & 26.4 & 0.262 \\
550 & 186.04470 & -1.09790 & ab & 17.77 & 17.69 & 26.8 & 0.997 \\
\enddata
\tablenotetext{a}{Angular distance from the center of 1220-1}



\end{deluxetable}

\begin{deluxetable}{ccccccccc}

\tablecaption{Oosterhoff characteristics of the LSQ survey and massive MW satellites}

\tablenum{5}

\tablehead{\colhead{Sample} & \colhead{$\langle P_{ab} \rangle$} & \colhead{$\sigma P_{ab}$} & \colhead{$\langle P_{c} \rangle$} & \colhead{$\sigma P_{c}$} & \colhead{$n_{ab}$} & \colhead{$n_{c}$} & \colhead{$n_{c}/(n_{ab} + n_{c})$} & \colhead{OoII \%} \\ 
\colhead{} & \colhead{(day)} & \colhead{(day)} & \colhead{(day)} & \colhead{(day)} & \colhead{} & \colhead{} & \colhead{} & \colhead{} } 

\startdata
All LSQ & 0.586 & 0.068 & 0.333 & 0.040 & 1007 & 358 & 0.26$\pm$0.02 & 27$\pm$2 \\
Sgr Stream & 0.591 & 0.062 & 0.337 & 0.038 & 194 & 70 & 0.27$\pm$0.04 & 32$\pm$5 \\
VSS Region & 0.574 & 0.063 & 0.336 & 0.035 & 73 & 20 & 0.22$\pm$0.05 & 24$\pm$7 \\
Smooth Halo & 0.589 & 0.069 & 0.321 & 0.034 & 160 & 62 & 0.28$\pm$0.04 & 25$\pm$5 \\
Smooth Halo \\
$R_{gc} < 20$ kpc & 0.585 & 0.069 & 0.312 & 0.033 & 64 & 28 & 0.30$\pm$0.07 & 29$\pm8$ \\
Smooth Halo \\
$30 < R_{gc} < 70$ kpc &  0.596 & 0.065 & 0.336 & 0.041 & 34 & 16 & 0.32$\pm$0.10 & 33$\pm$12 \\
\tableline
Sgr dSph &  0.575 & 0.064 & 0.326 & 0.034 & 1241 & 287 & 0.19$\pm$0.01 & \nodata \\
For dSph &  0.585 & 0.040 & 0.359 & 0.056 & 397 & 118 & 0.23$\pm$0.02 & \nodata \\
LMC & 0.576 & 0.073 & 0.329 & 0.041 & 17693 & 7213 & 0.29$\pm$0.00 & \nodata \\
SMC & 0.596 & 0.059 & 0.358 & 0.040 & 1933 & 542 & 0.22$\pm$0.01 & \nodata \\
\tableline
OoI GCs & 0.55 & \nodata & 0.32 & \nodata & \nodata & \nodata & 0.17 & \nodata \\
OoII GCs & 0.64 & \nodata & 0.37 & \nodata & \nodata & \nodata & 0.44 & \nodata \\
\enddata

\end{deluxetable}

\end{document}